\documentclass[aps,prd,twocolumn,showpacs,floatfix,preprintnumbers,amsmath,amssymb,nofootinbib]{revtex4}
\input epsf
\usepackage{graphicx}

\DeclareMathOperator{\atan}{atan}

\newcommand{\ket}[1]{\left| #1 \right>} 
\newcommand{\bra}[1]{\left< #1 \right|} 

\newcommand{\apjl}{Astrophys. J. Lett.}
\newcommand{\aap}{Astron. Astrophys.}
\newcommand{\apjs}{Astrophys. J. Suppl. Ser.}
\newcommand{\sa}{Sov. Astron. Lett.}
\newcommand{\jpb}{J. Phys. B.}
\newcommand{\natu}{Nature (London)}

\newcommand{\aj}{Astron. J.}
\newcommand{\aas}{Bull. Am. Astron. Soc.}
\newcommand{\mnras}{Mon. Not. R. Astron. Soc.}

\newcommand{\lsup}[2]{ \ensuremath{{}^{#2}\!{#1}}}
\newcommand{\lsub}[2]{ \ensuremath{{}_{#2}\!{#1}}}

\newcommand{\lsim}{\mathrel{\hbox{\rlap{\lower.55ex\hbox{$\sim$}} \kern-.3em \raise.4ex \hbox{$<$}}}}
\newcommand{\gsim}{\mathrel{\hbox{\rlap{\lower.55ex\hbox{$\sim$}} \kern-.3em \raise.4ex \hbox{$>$}}}}
\begin{document}
\title{Cosmological hydrogen recombination: The effect of extremely high-n states}
\author{Daniel Grin and Christopher M. Hirata} 
\affiliation{California Institute of Technology, Mail Code 350-17, Pasadena, CA 91125}
\date{\today}
\begin{abstract}
Calculations of cosmological hydrogen recombination are vital for the extraction of cosmological parameters from cosmic microwave background (CMB) observations, and for imposing constraints to inflation 
and reionization. The \textit{Planck} mission and future experiments will make high precision measurements of CMB anisotropies at angular scales as small as $\ell\sim 2500$, necessitating a calculation 
of recombination with fractional accuracy of $\approx 10^{-3}$. Recent work on recombination includes two-photon transitions from high excitation states and many radiative transfer effects. Modern 
recombination calculations separately follow angular momentum sublevels of the hydrogen atom to accurately treat nonequilibrium effects at late times ($z<900$). The inclusion of extremely high-n 
($n\gtrsim100$) states of hydrogen is then computationally challenging, preventing until now a determination of the maximum $n$ needed to predict CMB anisotropy spectra with sufficient accuracy for 
\textit{Planck}. Here, results from a new multi-level-atom code (\textsc{RecSparse}) are presented. For the first time, `forbidden' quadrupole transitions of hydrogen are included, but shown to be 
negligible. \textsc{RecSparse} is designed to quickly calculate recombination histories including extremely high-$n$ states in hydrogen. Histories for a sequence of values as high as $n_{\rm max}=250$ 
are computed, keeping track of all angular momentum sublevels and energy shells of the hydrogen atom separately. Use of an insufficiently high $n_{\rm max}$ value (e.g., $n_{\rm max}=64$) leads to errors 
(e.g., $1.8\sigma$ for \textit{Planck}) in the predicted CMB power spectrum. Extrapolating errors, the resulting CMB anisotropy spectra are converged to $\sim0.5\sigma$ at Fisher-matrix level for $n_{\rm 
max}=128$, in the purely radiative case.
\end{abstract}

\maketitle
\section{Introduction}
Measurements of cosmic microwave background (CMB) temperature anisotropies by the Wilkinson Microwave Anisotropy Probe (WMAP) have ushered in the era of precision cosmology, confirming that the Universe is spatially flat, with a matter budget dominated by dark matter and a baryonic mass fraction $\Omega_{b}h^{2}$ \cite{wmap_5} in agreement with the measured ratio of deuterium-hydrogen abundances (D/H) \cite{steigmanreview}. WMAP measurements of large-scale CMB polarization also yield the optical depth $\tau$ to the surface of last scattering (SLS), meaningfully constraining cosmological reionization. Together with surveys of supernovae \cite{perlmutter_supernovae,reiss_supernovae}, galaxies \cite{sdss_pspec,sdss_lrg,bao_discovery_cole}, and galaxy clusters \cite{cluster_cosmo}, WMAP measurements build the case that the Universe's expansion is accelerating, due to ``dark energy" or modifications of general relativity \cite{caldwell_darkenergy,carroll_1r}, and constrain other physical parameters (such as the sum of neutrino masses $\sum_{i}m_{\nu_{i}}$ \cite{neutrino,neutrinob,mabert} and the effective number of massless neutrino species $N_{\nu}$).

CMB temperature observations (WMAP, BOOMERANG \cite{boomerang_b}, CBI \cite{cbi} and ACBAR \cite{acbar_a}) probe properties of the primordial density field, such as the amplitude $A_{s}$, slope $n_{s}$, and running $\alpha_{s}$ of its power spectrum. These observations constrain deviations from the adiabatic, nearly scale free and Gaussian spectrum of perturbations predicted by the simplest models of inflation, but also offer controversial hints of deviations from these models (see Refs. \cite{wmap_5,acbar_c} and references therein).  Experimental upper limits to  B-mode polarization anisotropies (e.g. DASI \cite{dasi_a} and BICEP \cite{bicep}) impose constraints to the energy density of relic primordial gravitational waves \cite{seljak_pol,kamion_pol}.  

The \textit{Planck} satellite, launched in May 2009, will obtain extremely precise measurements of the CMB temperature anisotropy power spectrum ($C_{\ell}^{\rm TT}$) up to 
$\ell\sim 2500$ and the E-mode polarization anisotropy power spectrum ($C_{\ell}^{\rm EE}$) up to $\ell\sim 1500$ \cite{planck}. Robust measurements of the acoustic horizon and 
distance to the SLS will break degeneracies in dark energy surveys \cite{planck,detf_b,bao_discovery_eisenstein,bao_discovery_cole}. Polarization measurements will 
yield the optical depth $\tau$ to the SLS \cite{planck}, further constraining models of reionization and breaking the degeneracy between $n_{s}$ and $\tau$ 
\cite{planck}. Cosmological parameters will be determined with much greater precision. More precise values of $n_{s}$ and $\alpha_{s}$ will be obtained from CMB data 
alone, helping to robustly constrain inflationary models and alternatives to inflation \cite{planck}. The advent of \textit{Planck}, ongoing (SPT \cite{spt} and ACT 
\cite{act}) experiments at small scales, and a future space based polarization experiment like CMBPol \cite{cmbpol_c,cmbpol_d} all require predictions of primary 
anisotropy multipole moments $C_{\ell}$ with ${\cal O}(10^{-3})$ accuracy.

During atomic hydrogen (H) recombination, the Thomson scattering opacity drops, decoupling the baryon-photon plasma and freezing in acoustic oscillations. The phases of acoustic modes are set by the peak 
location of the visibility function \cite{bond_acoustic,peebles_acoustic}, while damping scales \cite{silk_damp,hu_sugiyama_silk_acoustic} and the amplitude of polarization \cite{bond_pol,polnarev_pol_a} are set by its width. Small-scale CMB anisotropies are also smeared out by free electrons along the line of sight, suppressing power on small scales so that $C_{\ell}\to C_{\ell}e^{-2\tau}$, where $\tau$ is the total optical depth of this w \cite{scottsmear}. An accurate prediction of the time-dependent free-electron fraction $x_{e}(z)$ from cosmological recombination is thus essential to accurately predict CMB anisotropies. 

Recent work has highlighted corrections of $\Delta x_{e}(z) /x_{e}(z)\gsim 0.1\%$ to the standard recombination history computed by \textsc{RecFast} \cite{seager_recfast}. These corrections will 
propagate through to predictions of anisotropies, and neglecting them would lead to biases and errors in \textit{Planck} measurements of cosmological parameters 
\cite{lewis_rec_cmb_param,wong_better_rec_b}. The use of the CMB as a 
probe of the first ionizing sources and of physics at energy scales greater than $10^{16}~{\rm GeV}$ thus requires an accurate treatment of  the $\sim {\rm eV}$ atomic physics of recombination \cite{improving_wong}.

Direct recombination to the hydrogen ground state is ineffective because of the high optical depth to photoionization \cite{peebles_rec,zeldovich_rec}. Recombination proceeds indirectly, first through recombination to a $n\geq2$ state of H, and then by cascades to the ground state. Because of the optical thickness of the Lyman-$n$ (Ly$n$) lines, the resulting radiation may be immediately absorbed, exciting atoms into easily ionized states. 

There are two ways around this bottleneck \cite{peebles_rec,zeldovich_rec}.  In the first, the sequence of decays from excited H levels ends with a two-photon decay (usually $2s\to 1s$). The emitted photons may have a continuous range of energies, allowing escape off resonance and a net recombination. In the second, 
photons emitted in the $np\to 1s$ transition redshift off resonance due to cosmological expansion, preventing re-excitation and yielding some net recombination. The 
dominant escape channel is from the $2p\to 1s$ Lyman-$\alpha$ line. These resonant transitions give off line radiation and distort the CMB 
\cite{rybicki_specdist_a,dubrovich_a}. 

Peebles, Sunyaev, Kurt, and Zel'dovich modeled recombination assuming that all net recombination resulted from escaping the $n=2$ bottleneck \cite{peebles_rec,zeldovich_rec}. This three-level-atom (TLA) treatment included recombinations to excited states, under the assumption of equilibrium between energy levels $n$ and angular momentum sublevels $l$ for all $n\geq 2$ (note the use of $l$ for atomic angular momentum and $\ell$ for CMB multipole number). This sufficed until the multi-level-atom (MLA) model of Seager et al. \cite{seager_phys_recfast}, which included hydrogen (H) and helium (He), separately evolved excited states assuming equilibrium between different $l$, accurately tracked the matter/radiation temperatures $T_{\rm M}$/$T_{\rm R}$ \cite{weymann_tmtr,sunyaev_tmtr}, accounted for line emission using the Sobolev approximation \cite{sobolev_sobolev_a}, and included ${\rm H}_{2}$ chemistry. This treatment underlies the \textsc{RecFast} module used by most CMB anisotropy codes, including those used for WMAP data analysis \cite{seager_recfast}.

The higher precision of \textit{Planck} requires new physical effects to be considered, among them two-photon transitions from higher excited states in H and He \cite{kholupenko_twophoton,chluba_twophoton_a,chluba_twophoton_b, hirata_twophoton,chluba_lymana_b}, other forbidden and semiforbidden transitions in He \cite{wong_scott_forbidden,hirata_helium_b,helium_forbidden}, feedback from Ly$n$ lines \cite{chluba_lyman_feedback}, and corrections to the Sobolev approximation due to a host of radiative transfer effects in H and He resonance lines \cite{kholupenko_helium_a,sunyaev_helium_a,chluba_lymana_a,chluba_lymana_b}. Most recent work on recombination has focused in one way or another on the radiative transfer problem. Here we direct our attention to the populations of very high-$n$ states.

One important effect is the breakdown of statistical equilibrium between states with the same value of the principal number $n$ but different angular momenta $l$. This effect is dramatic at late times. When $l$ sublevels of a level $n$ are resolved, increases in $x_{e}(z)$ of $\sim 1\%$ at late times result \cite{chluba_highn_a,chluba_highn_b}. This changes predicted $C_{\ell}$'s at a statistically significant level for \textit{Planck}. Highly excited states in hydrogen also change the recombination history at a level significant for \textit{Planck}. While levels as high as $n=300$ were included in the treatment of Ref.~\cite{seager_phys_recfast} underlying \textsc{RecFast}, $l$ sublevels were not resolved. It is thus important to update cosmological recombination histories to include high-n states of H while resolving $l$ sublevels, in order to predict the $C_{\ell}$'s as well as CMB spectral distortions from recombination.

Simultaneously including very high $n$ and resolving the $l$ sublevels is computationally expensive, taking nearly a week on a standard workstation for $n_{\rm max}=100$ \cite{chluba_highn_b}, using a conventional multilevel-atom recombination code. This becomes prohibitively expensive for higher values of $n_{\rm max}$, unless considerable resources are devoted to the problem. To date, this has prevented a determination of how $x_{e}(z)$ converges with $n_{\rm max}$ and how high $n_{\rm max}$ must be to predict $C_{\ell}$'s for \textit{Planck}. The existence of electric dipole selection rules $\Delta l=\pm 1$ means the relevant rate matrices are sparse, and we have used this fact to develop a fast code, \textsc{RecSparse}, to explore convergence with $n_{\rm max}$. While the computation time $t_{\rm comp}$ for standard $l$-resolving recombination codes scales as $t_{\rm comp} \propto n_{\rm max}^{6}$, with \textsc{RecSparse} the scaling is 
$t_{\rm comp}\propto n_{\rm max}^{\alpha}$, where $2<\alpha<3$. With \textsc{RecSparse}, we can calculate recombination histories for $n_{\rm max}=200$ in $~4$ days on a standard work-station; this would likely take weeks using a conventional code. For the first time, we have calculated recombination histories for $n_{\rm max}$ as high as $250$ \textit{with $l$ sublevels resolved}.

While previous computations have included some forbidden transitions, none have included optically thick electric quadrupole (E2) transitions in atomic hydrogen. We 
include E2 transitions, and find that the resulting correction to CMB anisotropies is negligible.

We find that the correction to CMB $C_{\ell}$'s due to extremely excited levels is $0.5\sigma$ or less if $n_{\rm max}\geq 128$, in the purely radiative case. This paper is not the final word on recombination; atomic collisions must be properly included and the effect of levels with $n>n_{\rm max}$ must be included to conclusively demonstrate absolute convergence. The end goal of the present recombination research program is to include all important effects in a replacement for \textsc{RecFast}, as the interplay of different effects is subtle.

In Sec. \ref{mla}, we review the formalism of the multilevel atom (MLA), and follow by explaining how we extend the MLA to include very high-n states (Sec. \ref{highnsec}) and electric quadrupole transitions (Sec. \ref{quadtheorysec}). State populations, recombination histories, and effects on the $C_{\ell}$'s are presented in Sec. \ref{res_highn}. We conclude in Sec. \ref{concl}.

We use the same fiducial cosmology as in Ref.~\cite{hirata_lymana}: total matter density parameter $\Omega_{m}h^{2}=0.13$, $\Omega_{b}h^{2}=0.022$, $T_{\rm CMB}=2.728~{\rm K}$, $N_{\nu}=3.04$,
and helium mass fraction $Y_{\rm He}=0.24$.
\section{The standard multilevel atom}
\label{mla}
We now review the elements of the standard multilevel-atom (MLA) treatment of cosmological recombination. For fundamental constants, we use NIST (National Institute of Standards and Technology) CODATA (Committee on Data for Science and Technology) values everywhere \cite{nist}. Unless explicitly noted otherwise, we make the substitution $m_{e}\to \mu=m_{e}m_{p}/\left(m_{e}+m_{p}\right)$ in all expressions for the Bohr radius $a_{0}$ and the ground-state hydrogen ionization potential $I_{\rm H}$ to correctly account for reduced-mass effects. 
\subsection{Basic framework}
CGS units are used except where explicitly noted otherwise. We follow the abundance $x_{n,l}=\eta_{n,l}/\eta_{\rm H}$, where $\eta_{\rm H}$ is the total number density of hydrogen nuclei and $\eta_{n,l}$ is the density of hydrogen in a state with principal quantum number $n$ and angular momentum $l$ (we denote the state $[n,l]$). We evolve these abundances including bound-bound and bound-free radiative, single photon, dipole transitions, as well as the $2s\to 1s$ two-photon transition, which has rate $\Lambda_{2s,1s}=8.2245809~{\rm s}^{-1}$ \cite{goldman}.
Focusing on the effect of single-photon dipole processes at high $n_{\rm max}$, we neglect higher $n$ two-photon processes  but note that their effects are large enough that they must be included in a final recombination code \cite{kholupenko_twophoton,chluba_twophoton_a,chluba_twophoton_b,hirata_twophoton,wong_scott_forbidden}. Note that we also neglect collisional transitions. We comment on how this may change our conclusions in Sec. \ref{lresec}.

Bound-bound electric dipole processes are described by the equation \cite{seager_phys_recfast,hirata_twophoton,peebles_rec}
\begin{eqnarray}
\left.\dot{x}_{n,l}\right|_{\rm bb}=\sum_{n^{\prime}\neq n,l^{\prime}=l\pm1}\left(\Gamma_{n,n^{\prime}}^{l ,l^{\prime}}x_{n^{\prime},l^{\prime}}-  \Gamma_{n^{\prime},n}^{l^{\prime},l}x_{n,l}\right),\label{mla_a}\end{eqnarray} with \begin{eqnarray}
\Gamma_{n,n^{\prime}}^{l,l^{\prime}}=\left\{\begin{array}{ll}
A_{n,n^{\prime}}^{l,l^{\prime}}P_{n,n^{\prime}}^{l,l^{\prime}}\left(1+\mathcal{N}^{+}_{nn^{\prime}}\right)&\mbox{if $n^{\prime}>n$,}\\ \\
A_{n^{\prime},n}^{l^{\prime},l}P_{n^{\prime},n}^{l^{\prime},l}\left(g_{l}/g_{l^{\prime}}\right)\mathcal{N}^{+}_{n^{\prime} n}&\mbox{if $n^{\prime}<n$},\label{mla_b}
\end{array}\right.
\end{eqnarray}
where $A_{n,n^{\prime}}^{l,l^{\prime}}$ is the downward Einstein rate coefficient for decays from $[n^{\prime},l^{\prime}]$ to $[n,l]$ and $P_{n,n^{\prime}}^{l,l^{\prime}}$ is the probability that a photon emitted in the $[n^{\prime},l^{\prime}]\to[n,l]$ line escapes the resonance without being reabsorbed. This probability is calculated in the Sobolev approximation, described in Sec. \ref{rad_trans}. For lower $l$ states easily described using the $s,p,d,f...$ orbital notation, we will sometimes use the notation $A_{1,n}^{0,1}=A_{np,1s}$, $P_{1,n}^{0,1}=P_{np,1s}$, and so on to simplify the discussion. The degeneracy of $[n,l]$ is $g_{l}=2(2l+1)$.  We explicitly keep track of the angular momentum quantum number $l$, as this will simplify discussion of our sparse-matrix technique in Sec. \ref{sparse_sec}.

The photon occupation number blueward/redward of a line transition  ($[n^{\prime},l^{\prime}]\to [n,l]$) is denoted
\begin{equation}
\mathcal{N}_{n n^{\prime}}^{\pm}=\mathcal{N}\left(E_{n,n^{\prime}}\pm \epsilon,T_{\rm R}\right),
\end{equation}
where $\mathcal{N}\left(E,T_{\rm R}\right)$ is the photon occupation number at photon energy $E$ and radiation temperature $T_{\rm R}$. Here $\epsilon$ is an infinitesimal line width and $E_{n,n\prime}$ is the energy of a photon produced in the transition $[n^{\prime},l^{\prime}]\to[n,l]$. The simplest possible assumption for $\mathcal{N}\left(E,T_{\rm R}\right)$ is a blackbody; we discuss further subtleties in Sec. \ref{rad_trans}:
\begin{equation}
\mathcal{N}\left(E_{n,n^{\prime}},T_{\rm R}\right)=\frac{1}{e^{E_{n,n^{\prime}}/\left(kT_{\rm R}\right)}-1}.
\end{equation} Here $k$ is the usual Boltzmann constant. The $\left(1+\mathcal{N}^{+}_{nn^{\prime}}\right)$ term accounts for stimulated and spontaneous emission.

The two-photon term is \cite{seager_phys_recfast,hirata_twophoton,peebles_rec}
\begin{eqnarray}
\left.\dot{x}_{2s\to 1s}\right|_{2\gamma}=-\left.\dot{x}_{1s\to 2s}\right|_{2\gamma}=\nonumber\\\Lambda_{2s\to 1s}\left[-x_{2s}+x_{1s}e^{-E_{2s,1s}/\left(kT_{\rm R}\right)}\right],\label{mla_c}
\end{eqnarray}
where $E_{2s,1s}=E_{2,1}$ and the second term describes two-photon absorption with a rate coefficient obtained by requiring that forward/backward rates satisfy the principle of detailed balance.

The bound-free term is \cite{seager_phys_recfast,hirata_twophoton,peebles_rec}
\begin{align}
&\left.\dot{x}_{n,l}\right|_{\rm bf}=\int \left[\eta_{\rm H}x_{e}^{2}\alpha_{nl}\left(E_{e}\right)S-x_{n,l}I\left(E_{e},T_{\rm r}\right)\right]dE_{e}\label{bfratenet},\end{align} with\begin{align}
&S\left(E_{e},T_{\rm M},T_{\rm R}\right)=\left[1+\mathcal{N}\left(E_{\gamma},T_{\rm R}\right)\right]P_{\rm M}\left(E_{e},T_{\rm M}\right)\end{align} and \begin{align}
&I\left(E_{e},T_{\rm R}\right)=\beta_{nl}\left(E_{e}\right)\mathcal{N}\left(E_{\gamma},T_{\rm R}\right).\label{mla_d}
\end{align}
This integral is over the total energy $E_{e}$ of a recombining electron. The energy of a recombination photon is $E_{\gamma}=E_{e}-E_{n}$, where $E_{n}$ is the bound-state energy of the recombined electron. The recombination rate in ${\rm cm}^{3}~{\rm s}^{-1}$ of such an electron to the bound state $\left[n,l\right]$ is $\alpha_{nl}\left(E_{e}\right)$ and is discussed in Sec. \ref{bfratesec}. The ionization rate in ${\rm s}^{-1}$ is $\beta_{nl}\left(E_{e}\right)$, and easily shown by detailed balance considerations to be \cite{hirata_twophoton}
\begin{equation}
\beta_{nl}\left(E_{e}\right)=\alpha_{nl}\left(E_{e}\right)\frac{2^{7/2}\pi\sqrt{E_{e}\mu^{3}}}{h^{3}g_{l}}.
\end{equation}

The free-electron abundance is $x_{e}=\eta_{e}/\eta_{\rm H}$, where $\eta_{e}$ is the free-electron density. We restrict our attention to times after helium recombination, and so the free proton abundance $x_{p}=x_{e}$. The net bound-free rate [Eq.~(\ref{bfratenet})] includes both spontaneous and stimulated recombination. The electron energy distribution is a Maxwellian with matter temperature $T_{\rm M}$:
\begin{equation}
P_{M}\left(T_{M},E_{e}\right)=2\sqrt{\frac{E_{e}}{\pi \left(kT_{\rm M}\right)^{3}}}e^{-E_{e}/\left(kT_{\rm M}\right)}.
\end{equation}
\subsection{Radiative transfer and escape probabilities}
\label{rad_trans}
Numerically solving the radiative transfer problem is computationally intensive, but tremendous simplification can be achieved with the Sobolev escape probability formalism, also known as the Sobolev approximation \cite{sobolev_sobolev_a}. The Hubble flow can be used to define a lengthscale over which the bulk flow induces a velocity change equal to the thermal velocity: $L=\sqrt{3kT_{\rm M}/m_{\rm atom}}/H(T_{\rm R})$, where $H\left(T_{\rm R}\right)$ is the value of the Hubble expansion parameter when the radiation has temperature $T_{\rm R}$ and $m_{\rm atom}$ is the mass of an atom \cite{seager_phys_recfast}. The conditions of the Sobolev approximation are \cite{hirata_lymana,seager_phys_recfast,sobolev_sobolev_a}: (i) $L$ is much smaller than the typical length scales over which cosmological quantities vary, (ii) $L/c$ is much smaller than the typical time scales over which cosmological quantities vary, (iii) complete frequency distribution--- the rest-frame frequency of an outgoing scattered photon $\nu$ does not depend on the incoming frequency $\nu^{\prime}$--- and (iv) no other emission, absorption, or 
scattering processes occur in the vicinity of the line. Corrections to the Sobolev approximation result from diffusion around resonance lines
\cite{krolik_lymana_a,rybicki_lymana_a}, atomic recoil \cite{grachev_lymana,hirata_lymana}, Thomson scattering near resonances \cite{chluba_lymana_plusthompson,hirata_helium_c}, and overlap of the higher Ly series lines, leading to important corrections to cosmological recombination calculations. In this work, however, we work in the Sobolev approximation to focus on other physical effects.

In the Sobolev approximation, the escape probability for photons produced in the downward transition $[n^{\prime},l^{\prime}]\to[n,l]$ is \cite{seager_phys_recfast} \begin{equation}
P_{n,n^{\prime}}^{l,l^{\prime}}=\frac{1-e^{-\tau_{n,n^{\prime}}^{l,l^{\prime}}}}{\tau_{n,n^{\prime}}^{l,l^{\prime}}},\end{equation} where the Sobolev optical depth is given by
\begin{equation}
\tau_{n,n^{\prime}}^{l,l^{\prime}}=\frac{c^{3}\eta_{\rm H}}{8\pi H\nu_{n,n^{\prime}}^{3}}A_{n,n^{\prime}}^{l, l^{\prime}}\left(\frac{g_{l^{\prime}}}{g_{l}}x_{n,l}-x_{n^{\prime},l^{\prime}}\right),\label{sobdepth}\end{equation} 
with transition frequency
\begin{equation}
\nu_{n,n^{\prime}}=\frac{E_{n,n^{\prime}}}{h}=\frac{{I_{\rm H}}}{h}\left|\frac{1}{n^{2}}-\frac{1}{n^{\prime2}}\right|.
\end{equation}
Correct expressions for $n^{\prime}<n$ are obtained by reversing arguments. During cosmological recombination, transitions between excited states are optically thin ($P_{n,n^{\prime}}^{l,l^{\prime}}\geq 0.99972$) \cite{hirata_twophoton}, and so we set $P_{n,n^{\prime}}^{l,l^{\prime}}=1$ in our calculations for non-Lyman lines. 

Transitions in the Lyman (Ly) series ($n^{\prime}>n=1$, $l^{\prime}=1$, $l=0$) are optically
thick ($\tau_{n,n^{\prime}}^{l,l^{\prime}}\gg 1$) \cite{hirata_twophoton}, and so $P_{1,n^{\prime}}^{0,1}\simeq 1/\tau_{1,n^{\prime}}^{0,1}$. Ly transitions cannot, however, be ignored in the recombination calculation, as the rate at which atoms find their way to the ground state through the redshifting of resonance photons, $P_{1,n^{\prime}}^{0,1}A_{1,n^{\prime}}^{0,1}$ is comparable to $\Lambda_{2s\to 1s}$ and other two-photon rates \cite{hirata_twophoton}. Strictly speaking, $\tau_{1,n^{\prime}}^{0,1}$ depends on $x_{n^{\prime},1}$, and so one should solve for $x_{n^{\prime} ,1}$ and then iteratively improve the solution. The populations of the excited states, however, are very small and the maximum resulting correction to the optical depth is $2\times10^{-12}$ (for $n^{\prime}=2,z=1600$) \cite{hirata_twophoton}. We thus drop the second term in Eq.~(\ref{sobdepth}), simplifying our computation by working in the approximation where the Lyman-$n$ (Ly$n$) line optical depth depends only  on the ground-state population and not on the excited-state populations.

Another aspect of the Lyman-series lines is feedback: a photon that escapes from the Ly$n$ ($np\rightarrow 1s$) line will redshift into the Ly$(n-1)$ line and be reabsorbed.  \textsc{RecSparse} has the 
ability to implement the resulting feedback, using the iterative technique of Ref. \cite{hirata_helium_a}.  This slows down the code by a factor of a few, however, and so to efficiently focus on the 
$n_{\rm max}$ problem, we turned feedback off.  For the high Lyman lines, feedback is almost instantaneous: the Universe expands by a factor of $\Delta\ln a\approx 2n^{-3}$ during the time it takes to 
redshift from Ly$n$ to Ly$(n-1)$.  In the instantaneous-feedback limit, the Ly$n$ lines do not lead to a net flux of H atoms to the ground state.  To approximate this net effect we turned off Lyman 
transitions with $n>3$; this leads to a smaller error than would result from leaving these transitions on
but disabling feedback.  Previous tests using the code of Ref.~\cite{hirata_lymana} show resulting errors in the recombination history at the $\approx1$\% level;
in any case, this should only weakly be related to the $n_{\rm max}$ problem.  All of the recombination histories and plots in this paper were produced by running
\textsc{RecSparse} with both feedback and Lyman transitions from $n>3$ disabled.

Electrons, though nonrelativistic during recombination, interact with photons through Thomson scattering. As a result, they do not follow the simple adiabatic scaling $T_{\rm M}\propto a^{-2}$, where $a$ is the cosmological scale factor. The matter temperature is set using the asymptotic solution of Ref.~\cite{hirata_twophoton} for $z>500$, after which the relevant ordinary differential equation (ODE) is solved numerically; this transition point occurs in the regime of mutual validity for the numerical and asymptotic solutions. We neglect subdominant processes, such as free-free, line, photorecombination and collisional ionization cooling, as well as photoionization and collisional recombination heating \cite{seager_phys_recfast}.

\subsection{The steady-state approximation}
The wide range of disparate time scales in this problem would naively necessitate a stiff differential equation solver. This computational expense can be avoided by repackaging Eqs.~(\ref{mla_a}),~(\ref{mla_b}),and ~(\ref{mla_c})-(\ref{mla_d}). These equations may be rewritten for excited states as ($[n,l]\neq[1,0]$)
\begin{eqnarray}
\dot{x}_{n,l}=-\sum_{n^{\prime}l^{\prime}} T^{l,l^{\prime}}_{n,n^{\prime}} x_{n^{\prime} ,l^{\prime}}+s_{n,l},\label{stiffset_a}\end{eqnarray} with \begin{eqnarray}
T^{l,l^{\prime}}_{n,n^{\prime}}=\delta_{n,n^{\prime}}^{l,l^{\prime}}\left(\mathcal{I}_{nl}+\gamma_{nl}+\sum_{n^{\prime\prime},l^{\prime \prime}}\Gamma_{n^{\prime\prime},n^{\prime}}^{l^{\prime\prime}, l^{\prime}}\right)-\Gamma_{n,n^{\prime}}^{l,l^{\prime}},\label{stiffset_b}\end{eqnarray} where
the integrated photoionization rate from $[n,l]$ is \begin{equation}
\mathcal{I}_{nl}=\int \beta_{nl}\left(E_{e}\right)I\left(E_{e},T_{\rm R}\right)dE_{e} \label{stiffset_c}\end{equation}
and $\Gamma_{n,n^{\prime}}^{l,l^{\prime}}$ is defined in Eq.~(\ref{mla_b}).

The downward flux to the ground state is \begin{equation}
\gamma_{nl}= A_{1,n}^{0,1}P_{1,n}^{0,1 }\left(1+\mathcal{N}_{1 n}^{+}\right)\delta_{l,1}+\Lambda_{2s,1s}\delta_{n,2 }^{l,0},\label{stiffset_d}
\end{equation}
where the first term describes Ly$n$ series transitions (stimulated and spontaneous) while the second accounts for the $[2,0]\to [1,0]$ two-photon transition. Kronecker delta symbols ($\delta_{n,n^{\prime}}^{l,l^{\prime}}$ and $\delta_{l,l^{\prime}}$) are employed throughout to enforce $[n,l]=[n^{\prime},l^{\prime}]$ and $l=l^{\prime}$).

The source term $s_{nl}$ includes flux from the ground state and direct recombination into the state $[n,l]$:
\begin{eqnarray}
s_{n,l}&=&\eta_{\rm H}x_{e}^{2}\int \alpha_{nl}\left(E_{e}\right)S\left(E_{e},T_{\rm M},T_{\rm R}\right)dE_{e}\nonumber\\
&+&x_{1s}\Lambda_{2s,1s}e^{-E_{2s,1s}/\left(kT_{\rm R}\right)}\delta_{n,2}^{l,0}\nonumber \\&+&x_{1s}g_{l}A_{1,n}^{0,1} P_{1,n}^{0,1} \mathcal{N}_{1 n}^{+}\delta_{l,1}/2.\label{stiffset_e}
\end{eqnarray}
This can also be rewritten in matrix notation: $d\vec{x}/dt=-\mathbf{T}\vec{x}+\vec{s}$, where $\mathbf{T}$ is the matrix of rates with components given by Eq.~(\ref{stiffset_b}).

The left-hand side of Eq.~(\ref{stiffset_a}) is associated with the recombination time scale, while both terms on the right-hand side are associated with much shorter atomic time scales. For example, the longest lifetimes in the recombination problem are those of the $2s$ and $2p$ states ($\Lambda_{2s,1s}\sim 10~{\rm s}$ and $A_{2p,1s}P_{2p,1s}\sim 1~{\rm s}$ when Ly-$\alpha$ optical depth peaks at $\tau\sim 6\times 10^{8}$), considerably shorter than the recombination time scale of $t_{\rm rec}\sim 10^{12}~{\rm s}$. Thus we make a steady-state approximation, $\dot{x}_{n,l}=0$, which is formally valid because the reciprocal of the minimum eigenvalue of $\mathbf{T}$ peaks at $0.8~{\rm s}$, which is $\sim10^{-12}$ of the duration of recombination. Thus the excited-state abundances are given by
\begin{equation}
\vec{x}\simeq\mathbf{T}^{-1}\vec{s}.
\end{equation}
The rates in $\mathbf{T}$ and $\vec{s}$ depend on $x_{e}$, $x_{1s}$, $T_{\rm M}$, $T_{\rm R}$, and $\mathcal{N}$. The ground-state population is given by $x_{1s}=1-x_{e}-\sum_{[n,l]\neq [1,0]}x_{n,l}$, but since excited-state populations are small ($x_{n,l}< 10^{-13}$), $x_{1s}$ can be eliminated from Eq.~(\ref{stiffset_e}) using the approximation $x_{1s}\simeq1-x_{e}$. We can then solve for the evolution of $x_{e}$, leaving out ineffective direct recombinations to the ground state:
\begin{eqnarray}
\begin{array}{l}
\dot{x}_{e}\simeq-\dot{x}_{1s}=x_{1s}\Lambda_{2s,1s}e^{-E_{2s,1s}/\left(kT_{\rm R}\right)}\\-\sum_{[n,l]\neq [1,0]}\left(\gamma_{nl}x_{n,l}-\frac{g_{l}}{2}A_{1,n}^{0,1}P_{1,n}^{0,1}\mathcal{N}_{1 n} x_{1s}\delta_{l,1}\right).\label{xe_evol}
\end{array}
\end{eqnarray}

The steady-state approximation thus allows us to convert a stiff system of ordinary differential equations into a large system of coupled linear algebraic equations, along with a single ordinary differential equation.
\section{Recombination with high-n states}
\label{highnsec}
The original ``effective 3-level atom" (TLA) treatments of cosmological recombination in Refs. \cite{peebles_rec,zeldovich_rec} were built on the assumption that the primary bottlenecks to effective recombination are the slow $2s\to 1s$ transition rate and the reabsorption of $2p\to 1s$ resonance photons by the optically thick plasma. Other crucial assumptions included radiative equilibrium between excited states,
\begin{eqnarray}
x_{n}=x_{2}e^{-\left(E_{n}-E_{2}\right)/\left(kT_{\rm R}\right)}n^{2}/4~~\mbox{if $n>2$},\\
x_{n}\equiv\sum_{l<n-1}x_{n,l},\label{rad_eq}\end{eqnarray} 
and statistical equilibrium between angular momentum sublevels:
\begin{equation}
x_{n,l}=x_{n}\frac{\left(2l+1\right)}{n^{2}}.\label{l_eq}
\end{equation}
Recombination to higher excited states was included through an effective ``Case B" total recombination constant $\alpha_{B}(T)$ (recombinations to the ground state are omitted) \cite{peebles_rec,seager_phys_recfast}.

As the radiation field cools and the baryon density falls at late times, the transitions coupling high-n to low-n become inefficient, as do those coupling different $l$ sublevels with the same $n$. This leads to a breakdown of statistical equilibrium (note however that the steady-state approximation is still valid), and so Eqs.~(\ref{rad_eq}) and (\ref{l_eq}) cease to apply. In Ref. \cite{seager_phys_recfast}, Eq.~(\ref{rad_eq}) is relaxed while Eq.~(\ref{l_eq}) is still imposed, and $\sim 10\%$ corrections to the TLA prediction for $x_{e}(z)$ result. At late times, nonequilibrium effects cause a net flux downward from states with quantum number $n$ to the ground state, accelerating recombination. 

The inclusion of progressively more shells increases the number of downward cascade channels to the ground state for continuum electrons. Thus higher $n_{\rm max}$
leads to faster recombination and lower $x_{e}(z)$. Reference \cite{seager_phys_recfast} reports results for $n_{\rm max}$ as high as $300$. The Lyman ($np\to 1s$)
transitions from very high-$n$ states overlap with the Lyman continuum, motivating Ref. \cite{seager_phys_recfast}'s claim that there is no need to go past $n=300$.
The real question as to whether the different values of $n$ are well defined, however, is whether the broadening of the state, $\hbar/\tau$ (where $\tau$ is the
lifetime) is larger than the splitting of adjacent energy levels, $\Delta E\approx 2I_{\rm H}n^{-3}$.  The intrinsic broadening for a typical level with $l/n\sim{\cal
O}(1)$ is $\hbar/\tau\sim\alpha^3I_{\rm H}n^{-5}$ \cite{brocklehurst_rates_a}.  Thus $\hbar/\tau\ll\Delta E$ and so these extremely high-$n$ energy levels are well
defined; indeed, transitions between highly excited states in such nonequilibrium plasmas are seen in interstellar H{\sc~ii} regions and are a useful diagnostic of
physical conditions \cite{goldberg_a}. 

For extremely large $n$, the above physical argument may break down because of additional broadening contributed by
interactions with the radiation field and the plasma.  For example, the broadening due to stimulated emission and absorption scales as $\sim n^{-2}$ (the spontaneous
$n^{-5}$ times the phase space density for photons in the $\Delta n=\pm1$ transitions) and that due to electron-impact collisions scales as $\sim n^{2}$
\cite{seaton_spec_a}; at sufficiently high $n$ these will dominate over $n^{-3}$ and the atomic energy levels will become blended.  However, the orders of magnitude
of the collisional coefficients \cite{seaton_spec_a} suggest that this occurs at values of $n$ larger than those considered in this paper. We have also verified that for conditions of interest for the recombining cosmological plasma, the plasma Debye length is greater than the average bound electron radius $a_{0}n^{2}$ as long as $n\lsim 10^{5}$. 

More recent work \cite{chluba_highn_a,chluba_highn_b} shows that additional $\sim1\%$ corrections to $x_{e}(z)$ arise when Eq.~(\ref{l_eq}) is not imposed and the populations of $l$ sublevels are followed separately. Bottlenecks to decays from high $l$ imposed by $l^{\prime}=l\pm 1$ slow down cascades to the ground state, and thus lead to slower recombination. In this case, the sidelength of $\mathbf{T}$ is $N=\mathcal{O}\left(n_{\rm max}^{2}\right)$. Since the number of computational steps needed to invert a matrix is generically a $N^{3}$ process, the computational time needed for a single ODE time step in the recombination time will be proportional to $n_{\rm max}^{6}$. 

As noted in Ref. \cite{chluba_highn_b}, a recombination calculation with $n_{\rm max}=100$ already takes $\sim 6$ days on a standard workstation. It this thus difficult to explore how quickly $x_{e}\left(z\right)$ converges for progressively higher values of $n_{\rm max}$. Even between $n_{\rm max}=80$ and $n_{\rm max}=100$, $\sim1\%$ changes are seen in the TT and EE multipole moments ($C_{\ell}$'s) of the CMB\footnote{In Ref. \cite{chluba_rico}, the results of Ref. \cite{chluba_highn_b} are used to explore the effect of progressively higher $n_{\rm max}$ on CMB $C_{\ell}$'s. In that work, It is noted that the fractional difference between the $C_{\ell}$'s for $n_{\rm max}=60$ and $n_{\rm max}=120$ falls within a heuristic \textit{Planck} performance benchmark. Higher values of $n_{\rm max}$ come even closer to the fiducial case of $n_{\rm max}=120$, a fact used to argue that even $n_{\rm max}=60$ recombination is adequate for \textit{Planck} data analysis. From the Cauchy convergence criterion, however, we know that a meaningful convergence test requires a comparison between successive members in a sequence. Using the results of Ref. \cite{chluba_highn_b} alone, the question of convergence with $n_{\rm max}$ thus remains open.} \cite{chluba_highn_b}. In spite of the computational challenge, it is thus crucial to push the calculation to sufficiently high $n_{\rm max}$ that corrections to $x_{e}(z)$ from remaining $n>n_{\rm max}$ are so small that they do not effect $C_{\ell}^{\rm TT}$ or $C_{\ell}^{\rm EE}$ at a level statistically significant compared to the predicted \textit{Planck} sample variance (e.g., several parts in $10^{4}$ for $l>1000$) \cite{hirata_helium_c}. There are two challenges in treating such a big multilevel atom. The first is the calculation of atomic transition rates at extremely high $n$; this is tractable because of some convenient recursion relations. The second is simultaneously evolving the populations of $n_{\rm max}\left(n_{\rm max}+1\right)/2$ states. We discuss these in turn below.

\subsection{Rates}
Here we discuss the Einstein coefficients for dipole bound-bound and bound-free transitions in atomic hydrogen, which are used in our recombination computation. We omit reduced-mass corrections to make a consistent comparison with Refs.~ \cite{brocklehurst_rates_a,hoangbinh_bb_numerical_rates,green_bb_numerical_rates,goldwire_bb_numerical_rates,burgess_bf_rates}, but include them when calculating actual recombination histories.

\subsubsection{Bound-bound rates}
The spontaneous electric dipole transition rate $\lsup{A_{n^{\prime},n}^{l^{\prime},l}}{\left(1\right)}$ for a nonrelativistic hydrogen atom is given by \cite{gordon_formula}
\begin{eqnarray}
\lsup{A_{n,n^{\prime}}^{l,l^{\prime}}}{\left(1\right)}=\frac{64\pi^{4}\nu_{n,n^{\prime}}^{3}}{3hc^{3}}\frac{{\rm max}(l,l^{\prime})}{2l+1}e^{2}a_{0}^{2}\left|\lsup{X_{n,n^{\prime}}^{l,l^{\prime}}}{\left(1\right)}\right|^{2},\label{aa}\\
\label{mel_bb}\lsup{X_{n,n^{\prime}}^{l,l^{\prime}}}{\left(1\right)}\equiv\left[\int_{0}^{\infty} x^{3}R_{n^{\prime}l^{\prime}}(x)R_{nl}(x)dx \right] ,\label{ab}
\end{eqnarray}
where $e$ is the charge of an electron, $h$ is the \textit{Planck} constant, and $\lsup{X_{n,n^{\prime}}^{l,l^{\prime}}}{\left(p\right)}$ denotes the radial matrix element between the states $\left[n,l\right]$ and $\left[n^{\prime},l^{\prime}\right]$ at order $p$ in the multipole expansion. For example, $\lsup{A_{n^{\prime},n}^{l^{\prime}, l}}{\left(2\right)}$ denotes the quadrupole rate, and so on.The restriction $l^{\prime}=l\pm 1$ enforces electric dipole selection rules.  Here $R_{nl}(x)$ is the radial wave function of an electron in a hydrogen atom, with principal quantum number $n$ and angular momentum quantum number $l$, at a dimensionless distance $x$. All dimensionless distances are measured in terms of $a_{0}$.
For Coulomb wave functions, this integration yields the Gordon formula \cite{gordon_formula}:
\begin{align}
&\lsup{X_{n,n^{\prime}}^{l,l^{\prime}}}{\left(1\right)}=\frac{\left(-1\right)^{n^{\prime}-l}}{4\left(2l-1\right)!} \sqrt{\frac{\left(n+l\right)!\left(n^{\prime}+l-1\right)!}{\left(n-l-1\right)!\left(n^{\prime}-l\right)!}}\label{hga}\\
&\times\frac{\left(4nn^{\prime}\right)^{l+1}}{\left(n+n^{\prime}\right)^{n+n^{\prime}}}\left(n-n^{\prime}\right)^{n+n^{\prime}-2l-2}W\left(n,n^{\prime},l\right)
,\nonumber\end{align}
where $l^{\prime}=l-1$, \begin{align}
W\left(n,n^{\prime},l\right)&=\lsub{F_{1}}{2}\left(u,-n^{\prime}+l,2l,w\right)-\left(\frac{n-n^{\prime}}{n+n^{\prime}}\right)^{2}\nonumber \\&\times \lsub{F_{1}}{2}\left(v,-n^{\prime}+l,2l,w\right),\end{align} with $u=-n+l+1$, $v=-n+l-1$, and $w=-4nn^{\prime}/\left(n^{\prime}-n\right)^{2}$. Here $\lsub{F_{1}}{2}\left(a,b,c;x\right)$ is Gauss's hypergeometric function for integer $a,b,$ and $c$, evaluated using the recursion relationship
\begin{align}
&\left(a-c\right)\lsub{F_{1}}{2}\left(a-1,b,c;x\right)=a(1-x)\left[\lsub{F_{1}}{2}\left(a,b,c;x\right)\right.\nonumber\\
&\left.-\lsub{F_{1}}{2}\left(a+1,b,c;x\right)\right]+\left(a+bx-c\right)\lsub{F_{1}}{2}\left(a,b,c;x\right),\end{align} with initial conditions
\begin{eqnarray}\lsub{F_{1}}{2}\left(0,b,c;x\right)=1,~~~\lsub{F_{1}}{2}\left(-1,b,c;x\right)=1-\frac{bx}{c}.\label{hgc}
\end{eqnarray}
We use Eqs.~(\ref{ab})-(\ref{hgc}) to calculate bound-bound transition rates at the beginning of a MLA computation, storing them for easy and repeated access. 

We compared the resulting radial matrix elements with several values for high $n$ in Ref.~ \cite{hoangbinh_bb_numerical_rates} and found agreement to all $3$ published digits. We calculated oscillator 
strengths and compared with Ref.~\cite{green_bb_numerical_rates} (all transitions with $n$ and $n'$ were evaluated, as was the entire Balmer series for $n\leq60$) and found agreement to all $6$ published digits. We also compared with the results in Ref.~ \cite{goldwire_bb_numerical_rates} (in which oscillator strengths were computed up to $n=500$ for $\Delta n\leq5$) and found agreement to $5$ digits. We attribute the difference in oscillator strengths to the fact that a polynomial expansion of $\lsub{F_{1}}{2}$ was used in Ref.~\cite{goldwire_bb_numerical_rates}, rather than the more stable recursion relationship. We also compared with the dipole one-photon rates used for the $n_{\rm max}=30$ MLA computation of Ref.~\cite{hirata_twophoton}. Most rates agreed to $7$ or more significant figures. Transition rates between $s$ and $p$ orbitals only agreed to $\sim5$ significant figures. We ran our MLA model using the rates of Ref.~\cite{hirata_twophoton} and verified that these small disagreements do not lead to any differences in $x_{e}\left(z\right)$ at the desired level of accuracy.  Given the high quantum numbers considered, it was important to verify that no numerical instability plagues our numerical implementation of these recursions.  We thus checked matrix elements computed using Eqs.~(\ref{hga})-(\ref{hgc}) against values estimated using the WKB approximation, as detailed in the Appendix.
\label{bbratesec}
\subsubsection{Bound-free rates}
Bound-free rates are evaluated using the same principle, but one of the two states used to evaluate matrix elements must be a continuum Coulomb wave function. The resulting matrix element is \cite{brocklehurst_rates_b}\begin{equation}
g_{n,\kappa}^{l,l^{\prime}}=\frac{1}{n^{2}}\int_{0}^{\infty} x^3 R_{nl}(x) F_{\kappa l^{\prime}}(x) dx,
\end{equation}
where $F_{\kappa l^{\prime}}$ is the continuum Coulomb wave function for a recombining photoelectron with angular momentum quantum number $l'$ and dimensionless energy $\kappa^{2}=E_{e}/{I_{\rm H}}=\frac{h\nu}{I_{\rm H}}-1/n^{2}$. The energy of the outgoing photon is $h\nu$. This integral may also be evaluated in terms of hypergeometric functions, which in turn yields a recursion relationship for $g_{n,\kappa}^{l,l^{\prime}}$ \cite{burgess_bf_rates}:
\begin{align}
G_{n,\kappa}^{l,l^{\prime}}&\equiv\frac{g_{n,\kappa}^{l,l^{\prime}}}{
\left(2n\right)^{l-n}\sqrt{\frac{\left(n+l\right)!}{\left(n-l-1\right)!}\prod_{s=0}^{l^{\prime}}\left(1+s^{2}\kappa^{2}\right)}}\nonumber, \\
G_{n,\kappa}^{l-2,l-1}&=\left[4\left(n^{2}-l^{2}\right)+l\left(2l-1\right)\left(1+n^{2}\kappa^{2}\right)\right]\nonumber
G_{n,\kappa}^{l-1,l}\\ &-4n^{2}\left(n^{2}-l^{2}\right)\left[1+\left(l+1\right)^{2}\kappa^{2}\right]G_{n,\kappa}^{l,l+1}\nonumber, \\
G_{n,\kappa}^{l-1,l-2}&=\left[4\left(n^{2}-l^{2}\right)+l\left(2l+1\right)\left(1+n^{2}\kappa^{2}\right)\right]\nonumber
G_{n,\kappa}^{l,l-1}\\ &-4n^{2}\left[n^{2}-\left(l+1\right)^{2}\right]\left(1+l^{2}\kappa^{2}\right)G_{n,\kappa}^{l+1,l}\label{befmel}
.\end{align}
The initial conditions of the recursion are \cite{burgess_bf_rates}
\begin{align}&G_{n,0}^{n-1,n}=\sqrt{\frac{\pi}{2}}\frac{8n}{\left(2n-1\right)!}\left(4n\right)^{n}e^{-2n}
\nonumber,\\
&G_{n,\kappa}^{n-1,n}=\frac{1}{\sqrt{1-e^{-\frac{2\pi}{\kappa}}}}
\nonumber\times \frac{e^{2n-2\kappa^{-1}\atan\left(n\kappa\right)}}{\left(1+n^{2}\kappa^{2}\right)^{n+2}} 
G_{n,0}^{n-1,n}\nonumber,\\
&G_{n,\kappa}^{n-2,n-1}={\left(2n-1\right)\left(1+n^{2}\kappa^{2}\right)}nG_{n,\kappa}^{n-1,n}\nonumber,\\
&G_{n,\kappa}^{n-1,n-2}=\left(\frac{1+n^{2}\kappa^{2}}{2n}\right)G_{n,\kappa}^{n-1,n}.
\end{align}

These matrix elements are tabulated at the beginning of each MLA run for all $l<n\leq n_{\rm max}$, and $10^{-25}\leq\kappa^{2}n^{2}\leq 4.96\times10^{8}$; this range of $\kappa$ is partitioned into $50$ logarithmically spaced bins, with each bin containing $11$ equally spaced $\kappa$ values. Bound-free matrix elements were compared with tabulated values for low $n$ in Ref.~\cite{burgess_bf_rates} and agreed to all $4$ listed digits. Matrix elements were also compared with those used in Ref.~ \cite{hirata_twophoton}; we found agreement to one part in $10^{7}$, aside from $s-p$ transitions, as already discussed. 

The recombination rate to $\left[n,l\right]$ as a function of energy is then
\begin{align}
\alpha_{nl}\left(E_{e}\right)=\frac{4\sqrt{\pi}\alpha^{4}a_{0}^{2}cI_{\rm H}^{3/2}}{3n^{2}\left(kT_{\rm M}\right)^{3/2}}
\sum_{l^{\prime}=l\pm1}{\rm max} \left\{l,l^{\prime}\right\}\Theta_{n,\kappa}^{l,l^{\prime}}, \label{recratenet} \end{align} with \begin{align}
\Theta_{n,\kappa}^{l,l^{\prime}}=\left(1+\frac{n^{2}E_{e}}{I_{\rm H}}\right)^{3}\left|g_{n,\kappa}^{l,l^{\prime}}\right|^{2}.
\end{align}

At each value of $T_{M}$, the tabulated matrix elements, Eqs.~(\ref{bfratenet}) and (\ref{recratenet})  are used to calculate thermally averaged recombination rates, using an $11$-point Newton-Cotes \cite{newton_cotes} formula for the integration and neglecting stimulated emission. Large bins are added into the integral until it has converged to a fractional precision of $5\times 10^{-15}$. We compared our values with integrated rates tabulated in Ref.~\cite{burgess_bf_rates} and found agreement to all $4$ listed digits. Comparing with the rates used in Ref.~\cite{hirata_twophoton}, we found agreement to one part in $10^{7}$, aside from s-p transitions. \\ \\ In Saha equilibrium, \begin{eqnarray}
\eta_{e}^{2}\alpha_{nl}\left(E_{e}\right)\left[1+\mathcal{N}\left(E_{\gamma},T_{\rm R}\right)\right]P_{\rm M}\left(E_{e},T_{\rm M}\right)\nonumber\\=\eta_{\rm H}x_{n,l}\mathcal{N}\left(E_{\gamma},T_{\rm R}\right)\beta_{nl}\left(E_{\gamma}\right),
\end{eqnarray}
and so by the principle of detailed balance,
\begin{align}
&\int dE_{e}\beta_{nl}\left(E_{\gamma}\right)=\frac{x_{e}^{2}\eta_{\rm H}}{x_{n,l}}\int dE_{e}\alpha_{nl}\left(E_{e}\right)\nonumber\\&\times\left.\frac{\left[1+\mathcal{N}\left(E_{\gamma},T_{\rm R} \right)\right]}{\mathcal{N}\left(E_{\gamma},T_{\rm R}\right)} \right|_{\rm eq} P_{\rm M}\left(E_{e}\right).
\end{align}
We verified that our computed thermally averaged recombination and ionization rates satisfied this equality to machine precision. We also checked bound-free matrix elements computed using Eq.~(\ref{befmel}) against values estimated using the WKB approximation, as detailed in the Appendix \ref{appendix_wkb}.
\label{bfratesec}
\subsection{Sparse-matrix technique}
\label{sparse_sec}

The key to making the recombination problem tractable for high values of $n_{\rm max}$ is the sparsity of Eqs.~(\ref{stiffset_a}) and (\ref{stiffset_b}). Dipole selection rules only allow coupling of states with angular momentum quantum numbers $l$ and $l^{\prime}$ if $l^{\prime}=l\pm 1$. It is easiest to understand how sparsity simplifies the problem with a slight change of notation. We can compose the vector $\vec{x}$ (with components $x_{n,l}$) of excited-state populations, as
\begin{equation}
\vec{x}=\left(\begin{array}{c}\vec{v}_{0}\\
\vec{v}_{1}\\
...\\
\vec{v}_{l_{\rm max}}
\end{array}\right),
\end{equation}
where $l_{\rm max}=n_{\rm max}-1$ and $\vec{v}_{l}$ denotes a vector of the populations of all states with angular momentum $l$, except for the $1s$ state. Specifically,
\begin{equation}
\vec{v}_{l}=\left(\begin{array}{c}x_{n_{\rm min},l}\\x_{n_{\rm min}+1,l}\\...\\
x_{n_{\rm max},l}\end{array}\right),
\end{equation}
where 
\begin{equation}
n_{\rm min}=\left\{\begin{array}{ll}2&\mbox{if $l=0$,}\\l+1&\mbox{if $l\neq 0$.}\end{array}\right.
\end{equation}
\begin{figure*}[ht]
\includegraphics[width=6.0in]{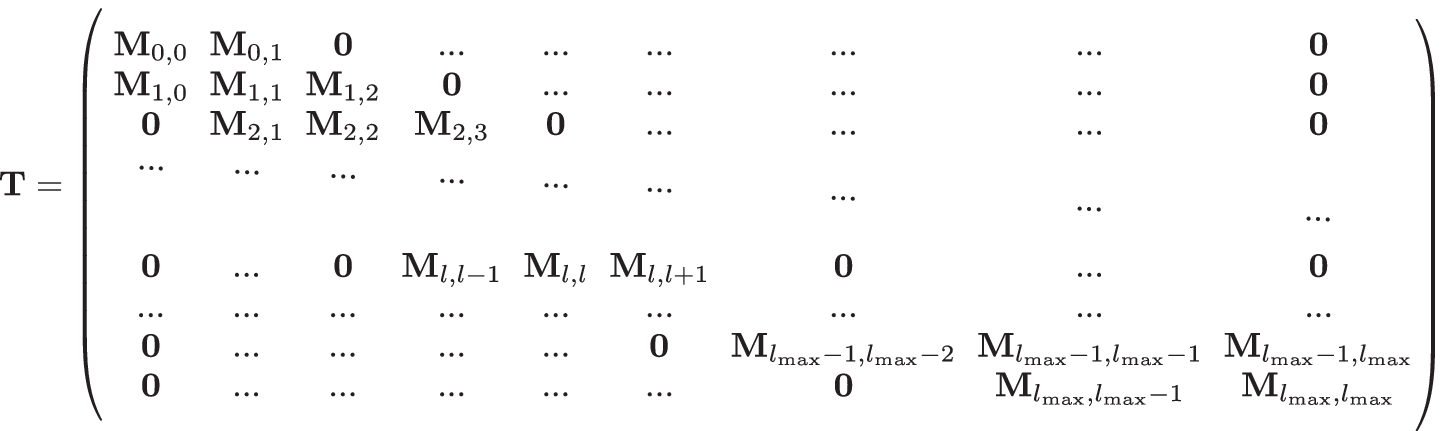}
\caption{Schematic of the sparse rate matrix $\mathbf{T}$ with components given by Eq.~(\ref{stiffset_b}) and submatrix building blocks given by Eq.~(\ref{submat_comp}). Boldface zeroes denote block matrices of all zeros, and enforce the dipole selection rule that the initial state $l^{\prime}$ angular momentum obeys $l^{\prime}=l\pm 1$, where $l$ is the final state angular momentum. The submatrix $\mathbf{M}_{l l^{\prime}}$ has dimension $\left(n_{\rm max}-n_{\rm min}+1\right)\times\left(n_{\rm max}-n_{\rm min}^{\prime}+1\right)$, where $n_{\rm min}=2$ if $l=0$, and $n_{\rm min}=l+1$ if $l>0$. Note that submatrices $\mathbf{M}_{l,l}$ on the block diagonal of the larger rate matrix $T$ are themselves diagonal, as seen from Eq.~(\ref{submat_comp}) and the fact that in the purely radiative case, $\Gamma_{n,n^{\prime}}^{l,l^{\prime}}=0$ if $n\neq n^{\prime}$ and $l= l^{\prime}$.}
\label{mfig}
\end{figure*}
The source vector $\vec{s}$ can similarly be written by concatenating source vectors $\vec{s}_{l}$; 
each $\vec{s}_{l}$ feeds all states with angular momentum $l$. 

The rate matrix may be similarly built of submatrices $\mathbf{M}_{l,l^{\prime}}$, as illustrated in Fig. \ref{mfig}. The complete rate matrix is block tridiagonal, and the blocks decrease in dimension as $l$ increases. 
The matrix $\mathbf{M}_{l,l^{\prime}}$ has components \begin{equation}M_{l,l^{\prime}}^{n,n^{\prime}}=\delta_{n,n^{\prime}}^{l,l^{\prime}}\left(\mathcal{I}_{nl}+\gamma_{nl}+\sum_{n^{\prime\prime},l^{\prime \prime}}\Gamma_{n^{\prime\prime},n^{\prime}}^{l^{\prime\prime}, l^{\prime}}\right)-\Gamma_{n,n^{\prime}}^{l,l^{\prime}}.\label{submat_comp}
\end{equation}

In the steady-state approximation, Eq.~(\ref{stiffset_a}) can be rewritten as a system of matrix equations. If  $l=0$,
\begin{equation}
\mathbf{M}_{0,0}\vec{v}_{0}+\mathbf{M}_{0,1}\vec{v}_{1}=\vec{s}_{0}.\label{bottom}
\end{equation}
If $0<l<l_{\rm max}$,
\begin{equation}
\mathbf{M}_{l,l-1}\vec{v}_{l-1}+\mathbf{M}_{l,l}\vec{v}_{l}+\mathbf{M}_{l,l+1}\vec{v}_{l+1}=\vec{s}_{l}.\label{mid}
\end{equation}
To close the system, we must truncate the hierarchy by excluding states with $n>n_{\rm max}$ as both sources and sinks, which is equivalent to setting $A_{n,n^{\prime}}^{l,l\pm 1}=0$ for ${\rm max}\left\{n,n^{\prime}\right\}>n_{\rm max}$. Then for $l=l_{\rm max}$, 
\begin{equation}
\mathbf{M}_{l_{\rm max},l_{\rm max}-1}\vec{v}_{l_{\rm max}-1}+\mathbf{M}_{l_{\rm max},l_{\rm max}}\vec{v}_{l_{\rm max}}=\vec{s}_{l_{\rm max}}.\label{top}
\end{equation} It might be possible to approximate the correction due to this truncation error, using asymptotic expressions for $A_{n,n^{\prime}}^{l,l\pm 1}$ and Saha equilibrium abundances for $n>n_{\rm max}$. This will only work if $n_{\rm max}$ is sufficiently high for nearly perfect equilibrium Saha equilibrium to hold between states with $n>n_{\rm max}$ and the continuum.

At any given time step, the actual quantity of interest is not the inverse $\mathbf{T}^{-1}$ of the rate matrix but the solution set $\left\{\vec{v}_{l}\right\}$ to the steady-state rate equations. The closed form solution to Eqs.~(\ref{bottom})-(\ref{top}) is
\begin{equation}
\vec{v}_{l}=\mathbf{K}_{l}\left[\vec{s}_{l}-\mathbf{M}_{l,l+1}\vec{v}_{l+1}+\sum_{l^{\prime}=0}^{l-1}\left(-1\right)^{l^{\prime}-l}\mathbf{S}_{l,l^{\prime}}\vec{s}_{l^{\prime}}\right], \label{ssola}
\end{equation}
if $l<l_{\rm max}$.
If $l=l_{\rm max}$, then
\begin{equation}
\vec{v}_{l}=\mathbf{K}_{l}\left[\vec{s}_{l}+\sum_{l^{\prime}=0}^{l-1}\left(-1\right)^{l^{\prime}-l}\mathbf{S}_{l,l^{\prime}}\vec{s}_{l^{\prime}}\right].\label{ssolb}
\end{equation}
Here
\begin{equation}
\mathbf{K}_{l}=\left\{\begin{array}{ll}\mathbf{M}_{00}^{-1}&\mbox{if $l=0$,}\\
\left(\mathbf{M}_{l,l}-\mathbf{M}_{l,l-1}\mathbf{K}_{l-1}\mathbf{M}_{l-1,l}\right)^{-1}
&\mbox{if $l>0$,}
\end{array}\right.\label{saa}
\end{equation}
and
\begin{equation}
\mathbf{S}_{l,i}=\left\{\begin{array}{ll}
\mathbf{M}_{l,l-1}\mathbf{K}_{l-1}&\mbox{if $i=l-1$,}\\
\mathbf{S}_{l,i+1}\mathbf{M}_{i+1,i}\mathbf{K}_{i}&\mbox{if $i<l-1$}.\end{array}\label{sab}
\right.
\end{equation}
Our new MLA code, \textsc{RecSparse}, operationally implements this solution at each time step as follows:
\begin{enumerate}
\item{Using the values of $T_{\rm R}$ and $x_{e}$, $T_{\rm M}$ is calculated using the results of Sec. \ref{rad_trans}.}
\item{All relevant $\mathbf{M}_{l,l^{\prime}}$ and $\vec{s}_{l}$ are computed using Eqs.~(\ref{submat_comp}) and (\ref{stiffset_e}) and stored.}
\item{All $\mathbf{K}_{l}$ and $\mathbf{S}_{l,i}$ are computed using Eqs.~(\ref{saa})-(\ref{sab}) and stored.}
\item{Equation (\ref{ssolb}) is applied to obtain the solution for $\vec{v}_{l_{\rm max}}$.}
\item{Equation (\ref{ssola}) is iterated to obtain the solutions for all $\vec{v}_{l}$.}
\end{enumerate}

The free-electron fraction $x_{e}$ is then evolved forward in time using $\left\{\vec{v}_{l}\right\}$ and Eq.~(\ref{xe_evol}). It would also be interesting to compute the cumulative spectral distortion emitted by the line and continuum processes responsible for recombination \cite{dubrovich_a,sunyaev_specdist,chluba_highn_a,chluba_highn_b}. This fractional perturbation of $10^{-7}$ to the blackbody intensity of the CMB could be detectable with future experiments and would offer a test both of our understanding of recombination and of new physics behind the surface of last scattering (e.g., time variation of fundamental constants, energy injection by decaying/annihilating dark matter) \cite{paddy_decaydm_cmb,chen_decaydm_cmb,spacetime_vary,slatyer_decaydm_cmb,zahn_friedmann_rec}. This and the development of a fast code  for \textit{\textit{Planck}} data analysis including all the relevant physical effects will be the subject of future work.

\subsection{Numerical methods}
\textsc{RecSparse} begins at $z=1606$, assuming Saha equilibrium to compute the initial value of $x_{e}$ and setting $T_{\rm M}$ as discussed in Sec. \ref{rad_trans}. Excited-state populations are obtained using the method of Sec. \ref{sparse_sec}. Submatrix inversions are implemented using the double precision routine DGESVX from the LAPACK library \cite{lapack}. Time evolution of $x_{e}(z)$ with Eq.~(\ref{xe_evol}) is implemented using the $5^{\rm th}$-order Runge-Kutta-Cash-Karp (RKCK) implementation in \textit{Numerical Recipes} \cite{nrbook}. The rapid time scale for return to Saha equilibrium introduces a stiff mode into the equations at early times, necessitating care in the choice of a stepsize for the integrator. We were able to achieve relative precision of $\epsilon\sim 10^{-8}$ by placing $59$ time steps at $z\geq1538$ and $250$ steps in the range $200\leq z\leq 1538$, partitioning each interval into equally sized steps in $\Delta \ln{a}$; relative errors were estimated by halving step size and comparing values of $x_{e}(z)$ at identical time steps. The computation time $t_{\rm comp}$ for \textsc{RecSparse} scales as $t_{\rm comp}\propto n_{\rm max}^{\alpha}$, where $2<\alpha<3$. This is an empirical estimate for the range of $n_{\rm max}$ that we have explored, and may not extend to higher $n_{\rm max}$ values. In contrast, for standard MLA codes, $t_{\rm comp}\propto n_{\rm max}^{6}$. We can calculate recombination histories for $n_{\rm max}=200$ in $~4$ days on a standard workstation; this would likely take weeks using a conventional MLA code. 

\label{num_meth}
\section{Extension to electric quadrupole transitions}
\label{quadtheorysec}
Early work on recombination highlighted the importance of forbidden transitions, as half of the hydrogen atoms in the Universe form by way of the $2s\to1s$ ``forbidden" transition \cite{peebles_rec,zeldovich_rec}. Recent work has included additional ``forbidden" transitions in the MLA treatment, namely, two-photon transitions ($ns\to 1s$ and $nd\to 1s$) in H \cite{kholupenko_twophoton,chluba_twophoton_a,chluba_twophoton_b,hirata_twophoton}, two-photon and spin-forbidden transitions in He~\cite{helium_forbidden,chluba_lymana_b,wong_scott_forbidden,hirata_helium_b}, as well as electric quadrupole (E2) transitions in He \cite{hirata_helium_a,hirata_helium_c}. 


Until this work, the impact of E2 transitions in H on recombination has not been considered, even though they are optically thick for transitions to/from the ground 
state. For optically thick lines, the overall transition rate is proportional to $A_{n n^{\prime}}^{l^{\prime} l}/\tau_{n n^{\prime}}^{l^{\prime} l}$. Since $\tau_{n 
n^{\prime}}^{l^{\prime} l} \propto A_{n n^{\prime}}^{l^{\prime} l}$, the overall transition rate is independent of the rate coefficient. Transitions such as electric 
quadrupoles, which seem ``weaker"  judging from rate coefficients alone, can thus be as important as ``stronger" transitions, like the Ly$n$ lines. For example, this 
is why the semiforbidden He{\sc~i} 591\AA\ line is important in cosmological recombination \cite{hirata_helium_a,hirata_helium_c}. We thus include E2 quadrupole 
transitions in our MLA computation 
to properly assess their relevance for cosmological recombination. M1 (magnetic dipole) transition rates in H are typically suppressed by an additional factor of $10^{7}-10^{8}$, and are thus negligible \cite{jitrik_bb_quad_rates}.
\subsection{Rates}
The electric quadrupole (E2) Einstein A-coefficient for transitions from states $\left[n,l\right]$ to states $ \left[n^{\prime},l^{\prime}\right]$ is \cite{johnson_quads}:
\begin{equation}
\lsup{A_{n^{\prime},n}^{l^{\prime},l}}{\left(2\right)}=\frac{\alpha\omega_{n,n^{\prime}}^{5}a_{0}^{4}}{15g_{a}c^{4}}\left|\bra{n l}|Q^{\left(2\right)}|\ket{n^{\prime}l^{\prime}}\right|^{2},
\end{equation}
where the quadrupole matrix element is
\begin{eqnarray}
\bra{n l}Q^{\left(2\right)}\ket{n^{\prime}l^{\prime}}=\bra{l}|C^{\left(2\right)}|\ket{l^{\prime}}\,\,^{\left(2\right)}X_{n^{\prime},n}^{l^{\prime},l}.
\end{eqnarray}

The matrix elements of the reduced angular tensor operator $C^{\left(2\right)}$ are given by
\begin{equation}
\bra{l}|C^{\left(2\right)}|\ket{l^{\prime}}=\left(-1\right)^{l}\sqrt{g_{l}g_{l^{\prime}}}
\left(\begin{array}{lll}l&2& l^{\prime}\\
0&0&0 \end{array}\right),
\end{equation}
where the last factor is the well-known Wigner-$3J$ symbol. This operator is defined as
\begin{eqnarray}
\bra{l}|C^{(k)}|\ket{l^{\prime}}&=&\left(-1\right)^{l-m}\left(\begin{array}{lll}l&~k&~l^{\prime}\\
-m&q&m^{\prime} \end{array}\right)^{-1}\nonumber \\  &\times& \sqrt{\frac{4\pi}{2k+1}}\bra{l m}Y_{kq}\left(\theta,\phi\right)\ket{l^{\prime} m^{\prime}}.
\end{eqnarray}

The dimensionless radial quadrupole integral is
\begin{equation}
\lsup{X_{n^{\prime},n}^{l^{\prime},l}}{\left(2\right)}=\int_{0}^{\infty} x^{4} R_{n^{\prime}l^{\prime}}(x)R_{nl}(x) dx.
\end{equation}
The radial matrix element for the $nd\rightarrow1s$ transition is a special case of Eq.~(B.13) of Ref. \cite{hey_bb_rates} with $n'=1$:
\begin{eqnarray}
\lsup{X_{1,n}^{0,2}}{\left(2\right)}=(-1)^{n-1}2^{6}n^{4}\left[\frac{\left(n+2\right)!}{\left(n-3\right)!}\right]^{1/2}\frac{\left(n-1\right)^{n-4}}{\left(n+1\right)^{n+4}}.\label{crap}
\end{eqnarray}


\subsection{Inclusion in multilevel atom code}
The obvious way to include quadrupole transitions into our MLA code would be to generalize Eq. ~(\ref{mid}) to include $\Delta l=\pm 2$ transitions:
\begin{equation}
\begin{array}{rr}
\mathbf{M}_{l,l+2}\mathbf{v}_{l+2}+\mathbf{M}_{l,l+1}\mathbf{v}_{l+1}+\mathbf{M}_{l,l} \mathbf{v}_{l}\\
+\mathbf{M}_{l,l-1}\mathbf{v}_{l-1}+\mathbf{M}_{l,l-2}\mathbf{v}_{l-2}=\mathbf{s}_{l}.
\end{array}
\end{equation}
The resulting system is obviously not as sparse as in the dipole case, and solving for all $\mathbf{v}_{l}$ would be computationally more expensive, slowing down the whole MLA code. Since the contribution from even the largest quadrupole rates may turn out to be small, we pursue a computationally less expensive approach.

Higher energy E2 transitions will proceed much faster than lower energy ones, since E2 rates scale as $\omega_{n n^{\prime}}^{5}$. In particular, transitions to and from the $1s$ ground state will 
dominate any other quadrupole contributions to the recombination problem, since
\begin{equation}
\frac{^{\left(2\right)}A_{1,n}^{0,2}}{^{\left(2\right)}A_{q,n}^{0,2}}\sim \frac{\omega_{1 n}^{5}}{\omega_{q n}^{5}}=\left[\frac{q^{2}\left(n^{2}-1\right)}{n^{2}-q^{2}}\right]^{5}\gsim10^{3}~\mbox{if $q\geq 2$}.
\end{equation}
Moreover, the $nd\rightarrow 1s$ lines are optically thick for small $n$.
We thus restrict our consideration to $nd\leftrightarrow1s$ transitions, since other quadrupole transitions are ``corrections to a correction." A further 
simplification follows if we recall that the Ly$n$ lines are all optically thick \cite{hirata_twophoton}. Thus, the transition $nd\to 1s$ is highly probable to be 
immediately followed by a transition $1s\to np$. This yields a net $nd\to np$ transition, analogous to an $l$-changing collision, which occurs with forward rate $\lsup{\Gamma^{0,2}_{1,n}}{\left(2\right)}=x_{nd}\lsup{A_{1,n}^{0,2}}{\left(2\right)}$. The reverse process occurs with rate $\lsup{\Gamma^{0,2}_{1,n}}{\left(2\right)}=x_{np}\lsup{A_{1,n}^{0,2}}{\left(2\right)}D$, where $D$ is a factor relating forward and backward rates.
If the $p$ and $d$ states were in equilibrium, the two rates would cancel, so by the principle of detailed balance, 
$D=\left(x_{nd}/x_{np}\right)_{\rm eq}=5/3$, where ``${\rm eq}$" denotes an equilibrium value. The net $np\leftrightarrow nd$ transition rate due to E2 transitions is thus
\begin{equation}
\dot{x}_{np}=-\dot{x}_{nd}=\lsup{A_{1,n}^{0,2}}{\left(2\right)}\left(x_{nd}-\frac{5}{3}x_{np}\right).\label{netquadrate}
\end{equation}
Since this overall rate obeys the $\Delta l=\pm 1$ selection rule, it can be numerically implemented within the same framework as the dipole rates.

\section{Results}
We ran the \textsc{RecSparse} code for a variety of $n_{\rm max}$ values. Here we omitted E2 transitions to focus on the effect of deviations from statistical equilibrium and increasing $n_{\rm max}$. We begin by discussing deviations from equilibrium, and proceed to discuss the recombination history and numerical convergence with $n_{\rm max}$. 
\label{res_highn}
\subsection{State of the gas}
\label{res_state_gas}
The assumptions of statistical equilibrium between different $l$ sublevels within the same $n$ shell and Boltzmann equilibrium between different $n$ states fail at late times, as discussed in Sec. \ref{highnsec}. Furthermore, as reactions become inefficient on the Hubble time scale and $x_{e}(z)$ freezes out, Saha equilibrium between the continuum and excited states of H also fails. Below, we discuss each of these failures quantitatively. 

\subsubsection{Populations of angular momentum sublevels}
\label{lresec}
At early times, the populations of hydrogen atoms in states with the same $n$ but different angular momentum $l$ are in statistical equilibrium [see 
Eq.$~$(\ref{l_eq})].  Radiative transitions do not include reactions that are $l$ changing but $n$ conserving. The $l$ sublevels must thus be kept in equilibrium by a 
combination of sequences of allowed radiative transitions and atomic collisions. These processes become inefficient at later times, leading the different 
$l$ sublevels to fall out of equilibrium. Both the TLA treatment of Peebles and the later MLA treatment of Seager et~al. rely on the statistical equilibrium assumption \cite{peebles_rec,seager_phys_recfast}. Our 
\textsc{RecSparse} code relaxes this assumption and follows the populations of all $l$ sublevels separately. 

\begin{figure}
\includegraphics[width=3.26in]{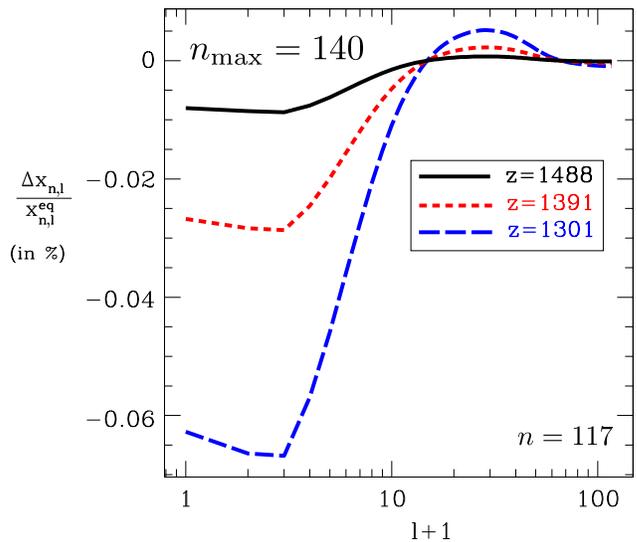}
\caption{Early time deviations from statistical equilibrium between different $l$ at fixed $n$ and $n_{\rm max}$, as computed by \textsc{RecSparse}.}
\label{res_state_gas_lneq_boltz_fine}
\end{figure}

\begin{figure*}
\includegraphics[width=5.7in]{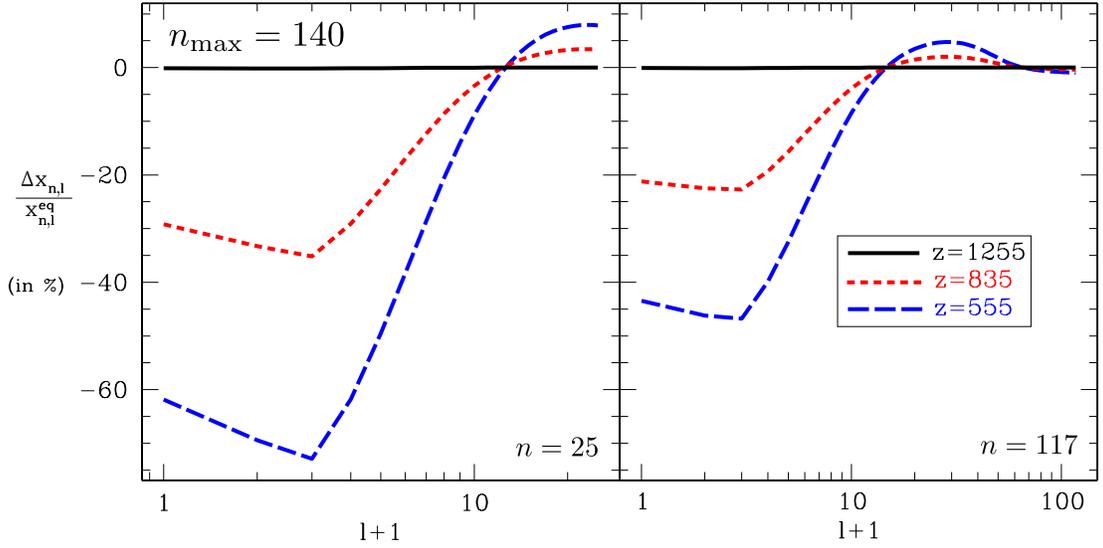}
\caption{Deviations from statistical equilibrium between different $l$ at fixed $n$ and $n_{\rm max}$, shown as computed by \textsc{RecSparse} at a variety of times through the recombination process. The left panel shows results for states with $n=25$, while the right panel shows results for states with $n=140$.}
\label{res_state_gas_lneq_boltz}
\end{figure*}

For $n>5$, the resulting populations are marked by several features, shown in Figs. 
\ref{res_state_gas_lneq_boltz_fine} and \ref{res_state_gas_lneq_boltz} at early and late times, respectively. We use \begin{equation}
\Delta x_{n,l}=x_{n,l}-x_{n,l}^{{\rm eq}}\end{equation} to compare actual with equilibrium populations, where \begin{equation}
x_{n,l}^{{\rm eq}}\equiv x_{n}\frac{\left(2l+1\right)}{n^{2}}.
\end{equation} Deviations begin modestly at early times ($\left|\Delta x_{n,l}/x_{n,l}^{\rm eq}\right|\lsim 0.1\%$ for $1300\lsim z\lsim 1600$) but are quite large by late times ($\left|\Delta x_{n,l}/x_{n,l}^{\rm eq}\right|\sim 60\%$ by $z\lsim 600$). 

Lower $l$ states depopulate efficiently, and are significantly underpopulated relative to statistical equilibrium expectations. States with $l=0$ can only make downward dipole transitions in $n$ if $l^{\prime}=1$. These rates are several order of magnitude lower than Lyman-series rates with the same $\Delta n$, and so $l=0$ states depopulate less efficiently than other low-$l$ states. This explains the upturn at the lowest $l$ values. The $\Delta l=\pm 1$ selection rule implies that higher $l$ states couple efficiently to neighboring bound states ($l^{\prime}=l\pm 1$) with a limited range of accessible $n^{\prime}$, since $n^{\prime}>l^{\prime}$. These states thus depopulate less efficiently than states with lower $l$ due to this bottleneck. 

The recombination rate $\alpha_{nl}$ peaks in the range $0.3\lsim l/l_{\rm max}\lsim 0.4$. Together, these facts imply the presence of a peak in $\Delta x_{n,l}/x_{n,l}^{\rm eq}$, which turns out to occur in the range $32\lsim l\lsim37$ for a wide range of $n$ at all times. The transition to $x_{n,l}/x_{n,l}^{\rm eq}\geq 1$ occurs in the range $16\lsim l\lsim 21$, also for a wide range of $n$ at all times. 
At very high $l$, recombination rates are so slow that these states are again underpopulated relative to statistical equilibrium, though less dramatically than they are at low $l$.

\begin{figure*}[ht]
\includegraphics[width=5.7in]{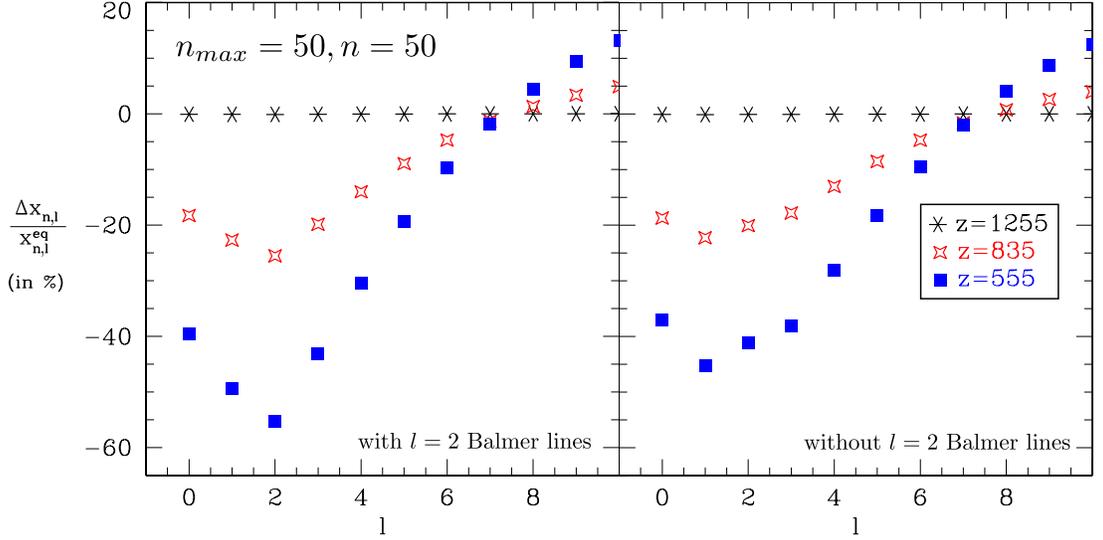}
\caption{The origin of the $l=2$ dip in Figs. \ref{res_state_gas_lneq_boltz_fine} and \ref{res_state_gas_lneq_boltz} is illustrated. Deviations from statistical equilibrium between different $l$ at fixed $n$ and $n_{\rm max}$ are shown at a variety of times through the recombination process. The left panel shows standard results with \textsc{RecSparse}. The right panel shows the results obtained if $l=2$ Balmer rates are artificially set to zero in the code. This figure highlights the relatively rapid $l=2$ Balmer transitions as the origin of the $l=2$ dip.}
\label{balmerfig}
\end{figure*}

The observed amplitude and shape of the curves in Figs. \ref{res_state_gas_lneq_boltz_fine}-\ref{res_state_gas_lneq_boltz} qualitatively agree with the results in Refs. 
\cite{chluba_highn_a}-\cite{chluba_highn_b}, including the upturn near the lowest $l$ and sharp minimum at $l=2$. The minimum is due to fast Balmer transitions out of the $l=2$ state. When we computed a 
recombination history with these rates ($nd\to2p$ for $n\geq 2$) artificially set to zero, the minimum moved to $l=1$, as shown in Fig. \ref{balmerfig}. It is interesting that the curves in Figs. 
\ref{res_state_gas_lneq_boltz_fine}-\ref{res_state_gas_lneq_boltz} exhibit the same behavior with $l$ as the departure coefficients of Ref. \cite{strel}, which describe neutral hydrogen (also in the 
steady-state approximation) in interstellar H{\sc~ii} regions.

\textsc{RecSparse} only takes into account radiative transitions, and omits $l$ and $n$-changing collisions. These rates would flatten all the curves in Figs. \ref{res_state_gas_lneq_boltz_fine}-\ref{balmerfig}, lessening deviations from statistical equilibrium between the different $l$ sublevels \cite{chluba_highn_b}. Indeed, the assumption of statistical equilibrium between these states at all times is formally equivalent to the limit of infinite $l$-changing collision rates. Theoretical estimates for collisional rates all depend on different assumptions and tabulated rates disagree by factors of two or more (see, e.g., Ref. \cite{clc}). As a function of redshift $z$, we estimate the ratio $f_{nl}^{\rm coll}\equiv \eta_{{\rm H}} x_{e} q_{nl}t_{nl}$ of collisional to radiative transition rates out of the state $\left[n,l\right]$, where $q_{nl}$ is the collisional rate coefficient (in ${\rm cm}^{3}~{\rm s}^{-1}$), and $t_{nl}$ is the total radiative lifetime of the state, including stimulated emission and absorption. 

Using the rate coefficients in Ref. \cite{pengelly}, we estimate that collisional rates (per unit time) are of the same order of magnitude as radiative rates for $n\gsim 52$ at $z \sim 1600$, $n\gsim 83$ at $z \sim 1080$, $n \gsim 160$ at $z\sim 740$, and $n\gsim 250$ at $z\sim 200$. In other words, as the primordial gas cools, collisions come to only influence the highest H energy levels, which contain the least bound electrons. This occurs because of the exponential decrease in the free-electron density $\eta_{{\rm H}}x_{e}$ in the early stages of recombination, which drives down collision rates accordingly. Near $z \sim 1600$ and shortly thereafter, radiative rates alone are high enough to keep the excited states in $l$-equilibrium. Collisions thus have little effect on $x_{e}(z)$ at early times. There may, however, be a window at some intermediate redshift, when collision rates are still relatively high, but departures from $l$-equilibrium are large enough to warrant including collisions in the recombination model. A full calculation is necessary to understand the actual effect. A final answer on the effect of resolving $l$ sublevels on both the recombination history $x_{e}(z)$ and the recombination spectrum awaits a robust theoretical calculation of the relevant collisional rates. This is an area of future investigation.

\subsubsection{Populations of Rydberg energy levels}
\begin{figure}[ht]
\includegraphics[width=3.26in]{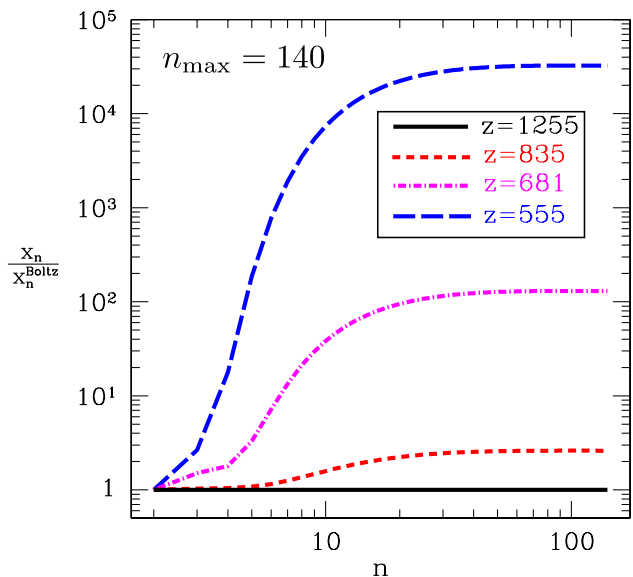}
\caption{Actual population of the $n^{\rm th}$ shell compared to its population in Boltzmann equilibrium with $n=2$, as computed by \textsc{RecSparse} at a variety of times through the recombination process.}
\label{res_state_gas_nneq_boltz}
\end{figure}

We may also compare the total population of the $n^{\rm th}$ energy level to values in Boltzmann equilibrium with $n=2$:
\begin{equation}
x_{n}^{\rm Boltz}\equiv x_{2}e^{-\left(E_{n}-E_{2}\right)/\left(kT_{R}\right)}n^{2}/4.\label{boldef}
\end{equation}
The recombination rate to states with $n>2$ is greater than the downward cascade rate, creating a bottleneck to depopulating these states. This bottleneck causes an over-population of the excited states compared to the equilibrium values of Eq.~(\ref{boldef}), as shown in Fig. \ref{res_state_gas_nneq_boltz}.
The ratio $x_{n}/x_{n}^{\rm Boltz}$ is $\mathcal{O}\left(1\right)$ at early times but grows as high as $3\times 10^{4}$ by $z=555$. The ratio approaches a constant at high $n$, as energy levels get closer to the continuum and the energy differences between successive levels shrink. 

Relative to $n=2$, excited states are over-populated, but there is no population inversion or cosmic maser. Excited states are still less populated than the $n=2$ energy level, just not as dramatically as they would be if Eq.~(\ref{boldef}) held. Among highly excited states, some pairs of levels do exhibit population inversion. For effective maser action, inversion must occur between pairs of radiatively connected levels, and the coherence of the radiation field must not be destroyed along the line of sight. 
This effect will be explored in detail in future work. In extremely dense structure-forming regions, more dramatic population inversion may result and lead to local masing; if these masers were observed, they could offer interesting new probes of structure formation near $z\sim1000$ as well as the physics of reionization \cite{spnorman}. 

\begin{figure}[ht]
\includegraphics[width=3.26in]{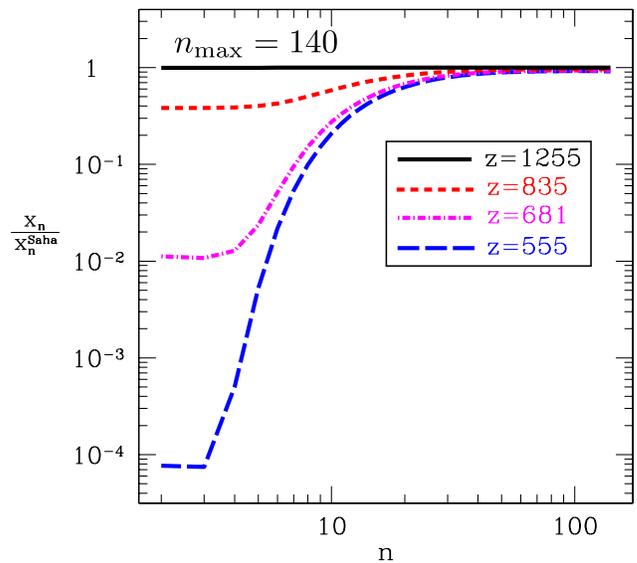}
\caption{Actual population of the $n^{\rm th}$ shell compared to the Saha equilibrium population, as computed by \textsc{RecSparse} at a variety of times through the recombination process.
}
\label{res_state_gas_nneq_saha}
\end{figure}

\begin{figure}[ht]
\includegraphics[width=3.26in]{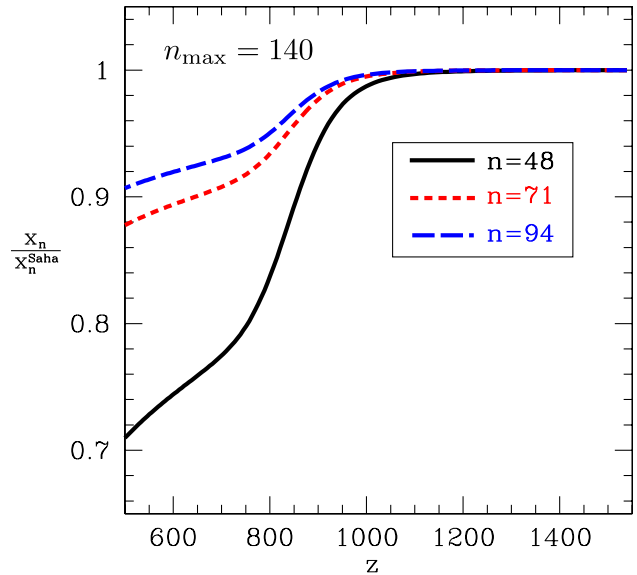}
\caption{Actual population of energy shells compared to Saha equilibrium values, shown for several $n$ values as an explicit function of cosmological redshift $z$.}
\label{res_state_gas_nneq_sahaz}
\end{figure}

Recombination becomes inefficient at late times; i.e., the recombination time scale $[\alpha_{\rm B}(T)x_e n_{\rm H}]^{-1}$ becomes longer than the age of the Universe. 
Saha equilibrium expressions for $x_{e}$ and $x_{\rm 1s}$ fail dramatically at late times. The free-electron fraction $x_{e}$ freezes out and is higher than the Saha equilibrium value, and thus $x_{\rm 1s}$ is lower than the Saha equilibrium value. Excited states are overpopulated relative to the ground state, but still not enough to be in Saha equilibrium with the continuum. The tower of excited states is thus also underpopulated relative to Saha equilibrium, as shown in Figs. 
\ref{res_state_gas_nneq_saha} and \ref{res_state_gas_nneq_sahaz}. Lower energy levels fall out of Saha equilibrium faster than higher energy levels. Higher energy 
levels are closest to Saha equilibrium, but at late times ($z\sim200$), even the population of the $n=250$ level is nearly $10\%$ below its Saha equilibrium value. 
Modeling the effect of states with $n>n_{\rm max}$ may require Saha equilibrium abundances to hold in the regime past the cutoff. To this end, it is important to 
properly model atomic collisions (which would push atoms towards Saha equilibrium at a lower transitional value of $n_{\rm max}$), and apply even greater 
computational resources to obtain $x_{e}(z)$ for even higher $n_{\rm max}$.

\subsection{The effect of extremely high-n states on recombination histories and the CMB}
\label{resultsn}

\begin{figure*}[ht]
\includegraphics[width=6.5in]{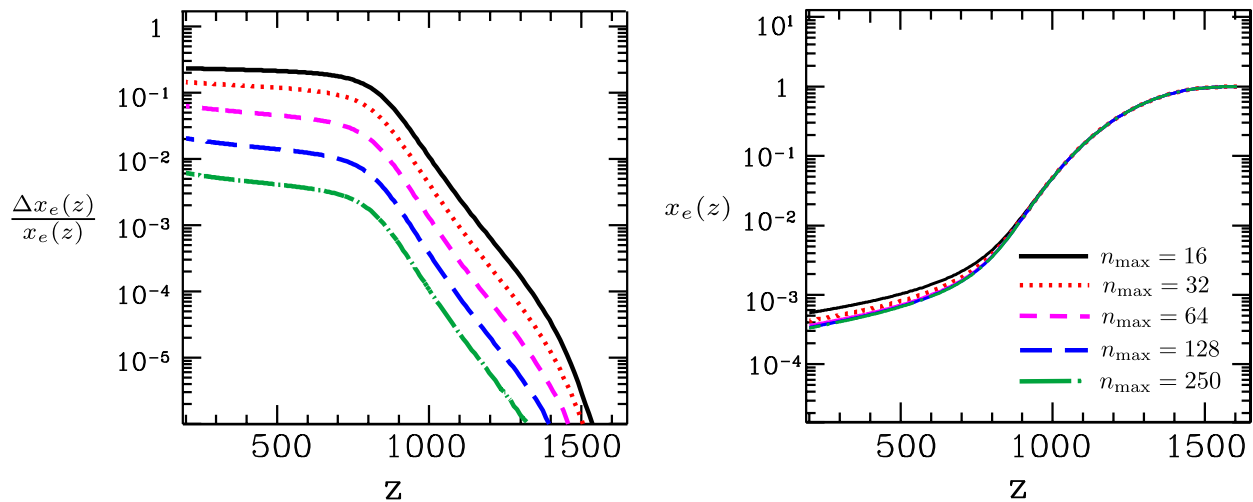}
\caption{The left panel shows relative errors between successively more accurate recombination histories with the indicated values of $n_{\rm max}$. Higher values of $n_{\rm max
}$ make recombination more efficient and yield lower freeze-out values of $x_{e}(z)$. As $n_{\rm max}$ increases, relative errors shrink, indicating that recombination is convergent with $n_{\rm max}$. The  right panel right panel contains the absolute recombination histories $x_{e}(z)$ and a legend. The relative error $\Delta x_{e}^{i}$ is defined in Eq.~(\ref{errdef}).}
\label{hist}
\end{figure*}

\begin{figure}[ht]
\includegraphics[width=3.26in]{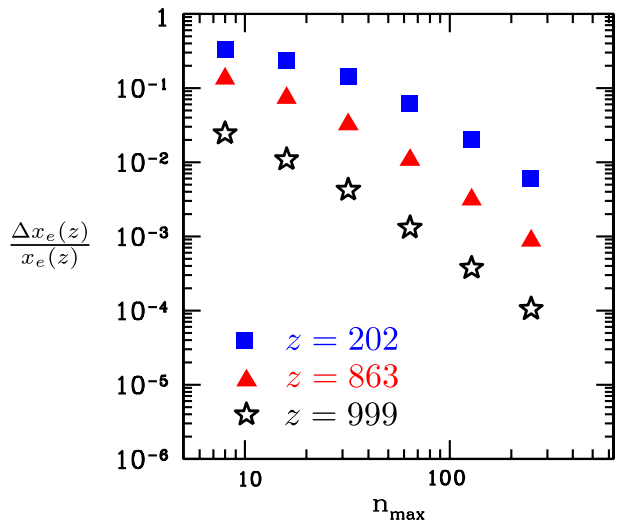}
\caption{Relative errors between successively more accurate recombination histories. Values are shown here for $3$ different values of redshift $z$. Errors shrink with $n_{\rm max}$, indicating relative convergence. Note, however, that this figure gives no scale for the absolute error.}
\label{conv}
\end{figure}

To explore the relative convergence of $x_{e}(z)$ over a wide logarithmic range of $n_{\rm max}$ values, we computed $x_{e}(z)$ for $n_{\rm max}=4,8,16,32,64,128$, and $250$. We define a relative error:
\begin{equation}
\Delta x_{e}^{i}\left(z\right)=x_{e}^{n_{\rm max}^{i-1}}\left(z\right)-x_{e}^{n_{\rm max}^{i}}\left(z\right).\label{errdef}
\end{equation}
Here $n_{\rm max}^{i}$ is the $i^{\rm th}$ $n_{\rm max}$ value. We show the resulting recombination histories and $\Delta x_{e}^{i}\left(z\right)$ in Fig. \ref{hist}. As $n_{\rm max}$ increases, the larger number of pathways to the ground state makes recombination more efficient, decreasing $x_{e}^{n_{\rm max}^{i}}(z)$ and making $\Delta x_{e}^{i}\left(z\right)$ positive. The relative error $\Delta x_{e}^{i}\left(z\right)$ shrinks with $n_{\rm max}$, indicating that relative convergence is taking place, as demonstrated in Fig. \ref{conv}. Note, however, that the relative error may not be a good proxy for the absolute error. Suppose that the absolute error is given by $x_{e}^{n_{\rm max}^{i}}=\Delta x_{e}^{{\rm abs},i}+x_{e}$, where $\Delta x_{e}^{{\rm abs},i}=A\left(n_{\rm max}^{i}\right)^{p}$, for some normalization $A$ and power-law index $p<0$. Then it is easy to show that for $n_{\rm max}^{i}=2n_{\rm max}^{i-1}$, $\Delta x_{e}^{i}/\Delta x_{e}^{\rm abs,i}=(1-2^{p})$. In other words, the relative error will underestimate the absolute error. To demonstrate absolute convergence, one should demonstrate that the physics neglected by ignoring transitions to $n>n_{\rm max}$ does not cause large changes in $x_{e}(z)$. We also calculated recombination histories for $n_{\rm max}=20,50,90,105$, and $160$.

\begin{figure}[ht]
\includegraphics[width=3.26in]{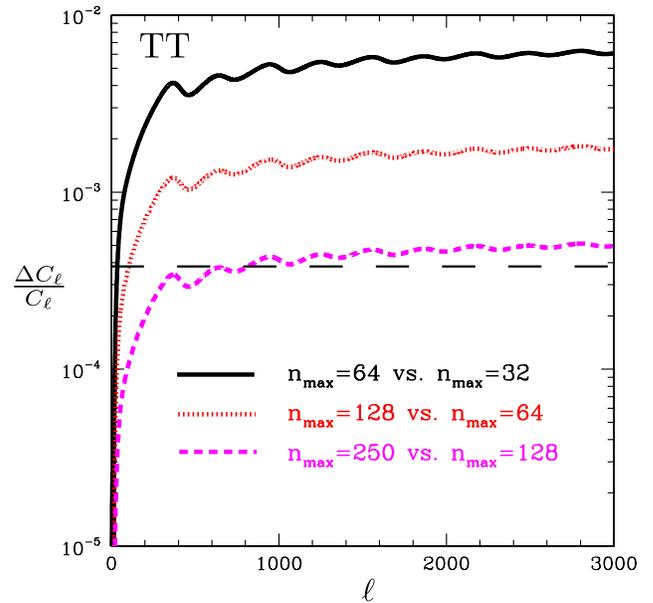}
\caption{Relative errors between temperature anisotropy spectra $C_{\ell}^{\rm TT}$ computed using \textsc{CMBFast}, modified to include successively more accurate \textsc{RecSparse} recombination histories. Pairs of $n_{\rm max}$ values used for the comparison are indicated in the legend. $C_{\ell}^{\rm TT}$ increases with $n_{\rm max}$, as discussed in Sec. \ref{resultsn}. The correction shrinks with increasing $n_{\rm max}$. The long dashed line indicates the cosmic variance target for $\Delta C_{\ell}/C_{\ell}$, as discussed in the text.}
\label{tt}
\end{figure}

\begin{figure}[ht]
\includegraphics[width=3.26in]{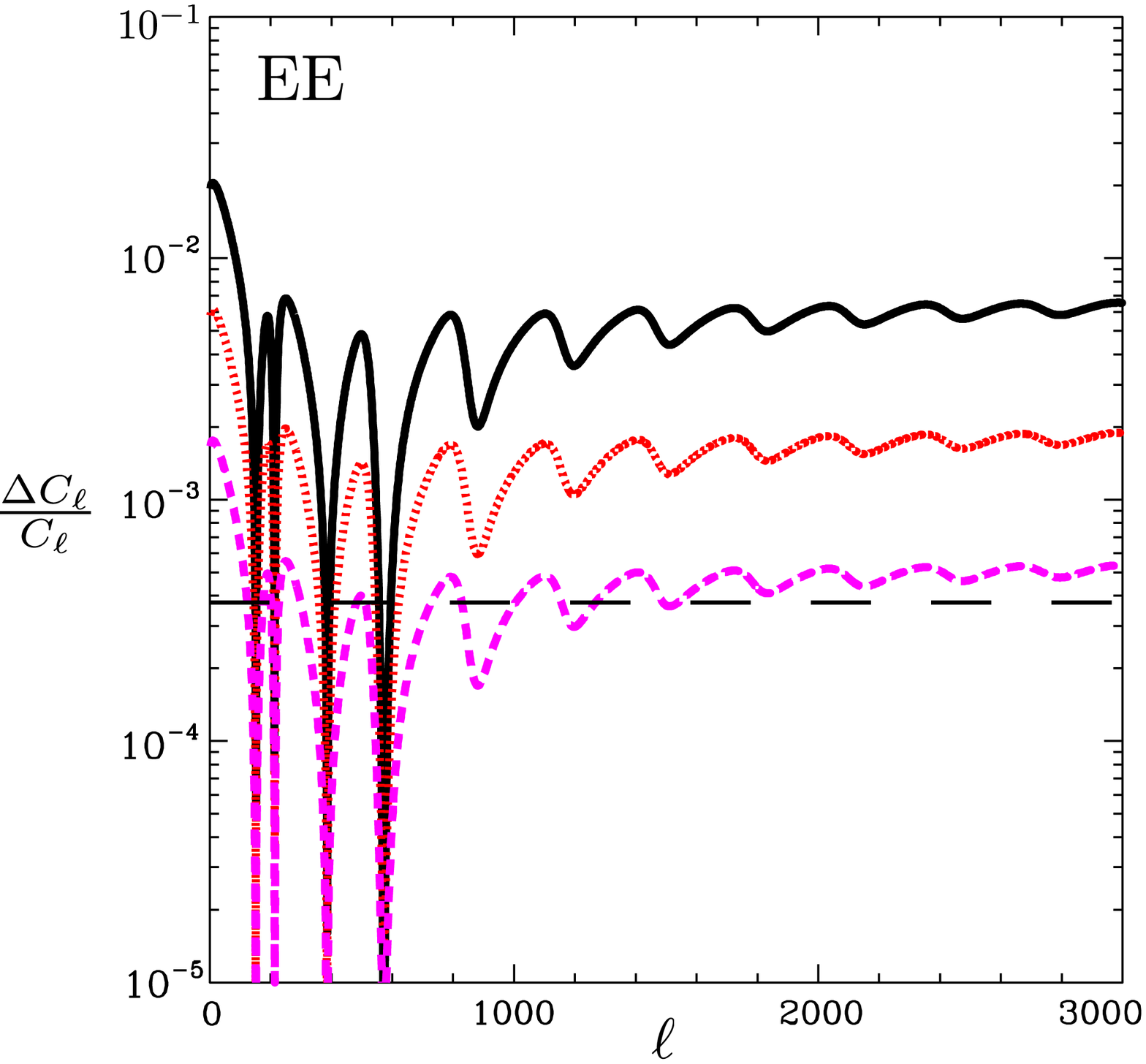}
\caption{Relative errors between E-mode polarization anisotropy spectra $C_{\ell}^{\rm EE}$ computed using \textsc{CMBFast}, modified to include successively more accurate \textsc{RecSparse} recombination histories. Pairs of $n_{\rm max}$ values used for the comparison are indicated in the legend of Fig. \ref{tt}. $C_{\ell}^{\rm EE}$ increases with $n_{\rm max}$, as discussed in Sec. \ref{resultsn}. The correction shrinks with increasing $n_{\rm max}$. The long dashed line indicates the cosmic variance target for $\Delta C_{\ell}/C_{\ell}$, as discussed in the text.}
\label{ee}
\end{figure}

We may also assess the effect of the computed changes in $x_{e}(z)$ on the CMB $C_{\ell}$'s.  To this end, we replace the usual table generated and used in the \textsc{RecFast} module of \textsc{CMBFast} with a table of our own output for different $n_{\rm max}$ values, smoothly stitching our history onto the usual \textsc{RecFast} history at the boundaries $z=1606$ and $z=200$. We tried a variety of smoothing schemes including no smoothing at all, and determined that the resulting error was at most $10\%$ the change already induced by varying $n_{\rm max}$. The choice of smoothing scheme is thus a ``correction to a correction" and does not alter the conclusions of our analysis. In particular, the number of sigmas at which power spectra corrected and uncorrected for higher-$n$ levels can be distinguished will change by at most $10\%$ of itself as a result of changing the smoothing scheme. The statistical significance of higher-$n$ shells will thus be essentially unchanged by the choice of smoothing scheme. The results for temperature and E-mode polarization anisotropy power spectra ($C_{\ell}^{\rm TT}$ and $C_{\ell}^{\rm EE}$) are shown in Figs. \ref{tt} and \ref{ee}, respectively. Here we also define a relative error:
\begin{equation}
\Delta C_{\ell}^{{\rm XX},i}=C_{\ell}^{XX,n_{\rm max}^{i-1}}-C_{\ell}^{{\rm XX},n_{\rm max}^{i}}.\label{errdef_microwave}
\end{equation}
Here XX denotes the TT or EE label of the power spectrum under consideration. The relative error $\Delta C_{\ell}^{\rm XX,i}$ is always positive, indicating that increasing $n_{\rm max}$ also increases $C_{\ell}^{XX}$, as shown in Figs. \ref{tt} and \ref{ee}. The common (TT and EE) origin for this effect is clear from Fig. \ref{hist}. Higher $n_{\rm max}$ makes recombination more efficient, driving down the freeze-out value of $x_{e}(z)$ and the residual optical depth $\tau$, leading to the high-$l$ plateaus seen in Fig. \ref{tt} and \ref{ee}. As a result, the smearing out of primary CMB anisotropies by relic free electrons, $C_{\ell}\to C_{\ell}e^{-2\tau}$ \cite{scottsmear}, is less dramatic when $n_{\rm max}$ is increased. The relative error $\Delta C_{\ell}^{\rm XX,i}$ shrinks with increasing $n_{\rm max}$. 

Taken as a proxy for the absolute error, $\Delta C_{\ell}^{\rm XX,i}$ may be compared to a crude (cosmic variance) estimate of the required accuracy of $C_{\ell}^{\rm XX}$ predictions in the damping tail:
\begin{equation}
\frac{\Delta C_{\ell}^{\rm XX}}{C_{\ell}^{\rm XX}}\sim 3\times 10^{-4} f_{\rm sky}^{-1/2}.\label{plancktarget}
\end{equation}

Here $f_{\rm sky}$ is the fraction of the sky covered by a CMB experiment. For $f_{\rm sky}=0.70$, results are shown in Figs. \ref{tt} and \ref{ee} and we see that only for $n_{\rm max}=250$ does the 
relative error shrink to a level comparable with the cosmic variance. The ultimate aim is for \textit{the total} correction from recombination physics to be less than statistical errors, so any 
individual contribution such as the truncation error at $n_{\rm max}$ should be $\ll1\sigma$.  In any case, collisions must be properly included to show absolute convergence, and so this should be a key 
focus of future work on highly excited states in hydrogen recombination. To more realistically assess the importance of high-$n$ states, $\Delta C_{\ell}^{XX}$ should be compared with a realistic error 
estimate for \textit{Planck}.

\subsection{Statistical significance of corrections to recombination history}
As a test of the importance of the modified recombination history for {\slshape Planck}, we have compared our corrections to the power spectrum $\Delta C_\ell$ with the 
forecast {\slshape Planck} error bars.  The comparison is done by means of the statistic
\begin{equation}
Z = \sqrt{\sum_{ll'} F_{ll'} \Delta C_l \Delta C_{l'}},
\end{equation}
where $F_{ll'}$ is the Fisher matrix for the CMB power spectrum.  For the temperature-only case, $\ell$ ranges from 2 to $\ell_{\rm max}$ and hence ${\bf F}$ is an $(\ell_{\rm 
max}-1)\times(\ell_{\rm max}-1)$ matrix; when polarization is included, ${\bf F}$ expands to a $3(\ell_{\rm max}-1)\times3(\ell_{\rm max}-1)$ matrix incorporating TT, EE, and 
TE spectra.  The $Z$ statistic is the number of sigmas at which the corrected and uncorrected power spectra could be distinguished assuming perfect knowledge of the 
cosmological parameters, and hence represents the largest possible bias (in sigmas) on any combination of cosmological parameters in any fit that incorporates the 
CMB \cite{hirata_twophoton}.  We use the forecast noise and beam curves for {\slshape Planck} data 70 GHz (Low-Frequency Instrument) and 100 and 143 GHz 
(High-Frequency Instrument) channels in the Blue Book \cite{planck}, and assume a usable sky fraction of $f_{\rm sky}=0.7$.

The computation considering the difference between the $n_{\rm max}=128$ and 250 curves gives a $Z$ value of $0.36$.  However, the actual error in the $n_{\rm 
max}=128$ calculation is somewhat greater because even the $n_{\rm max}=250$ calculation is not completely converged.  If the error in the $C_l$s scales as $\sim 
n_{\rm max}^p$ and has a shape that varies slowly with $n_{\rm max}$, then our value of $Z$ should be increased by a factor of $[1-(250/128)^p]^{-1}$; for $p\approx
-1.9$ (as suggested by Fig.~\ref{conv}) this is 1.39.  
Thus if the power-law extrapolation is to be trusted there is a $0.50\sigma$ error ($Z=0.50$) in the CMB power spectrum if one restricts attention to $n_{\rm 
max}=128$, and a $\sim 4$ times smaller error ($Z=0.14$) at $n_{\rm max}=250$.  A similar comparison between $n_{\rm max}=64$ and 250 implies an error of 
$Z=1.79$ at $n_{\rm max}=64$.  This suggests that in the {\em purely radiative} problem the CMB power spectrum is converged (in the sense that our remaining errors are small compared to 
projected {\slshape Planck} errors) at $n_{\rm max}\ge128$; however this issue will have to be reconsidered in future work when collisions are included.

\subsection{The effect of electric quadrupole transitions on recombination histories and the CMB}
\label{res_quad}
Using the treatment of Sec. \ref{quadtheorysec} and an integration stepsize fine enough to obtain a fractional accuracy of $10^{-10}$ in $x_{e}$, we compute the effect of E2 quadrupole transitions on cosmological hydrogen recombination for several values of $n_{\rm max}$. We can parametrize this effect using 
\begin{equation}
\Delta x_{e}\equiv \left.x_{e}\right|_{{\rm no}~E2~{\rm transitions}}-\left.x_{e}\right|_{{\rm with}~E2~{\rm transitions}} 
\end{equation}
and 
\begin{equation}
\Delta C_{\ell}\equiv \left.C_{\ell}\right|_{{\rm with}~E2~{\rm transitions}}-\left.C_{\ell}\right|_{{\rm no}~E2~{\rm transitions}}.\end{equation} Note that unlike the case of varying $n_{\rm max}$, these are the absolute errors induced by ignoring E2 transitions.

\begin{figure}[t]
\includegraphics[width=3.26in]{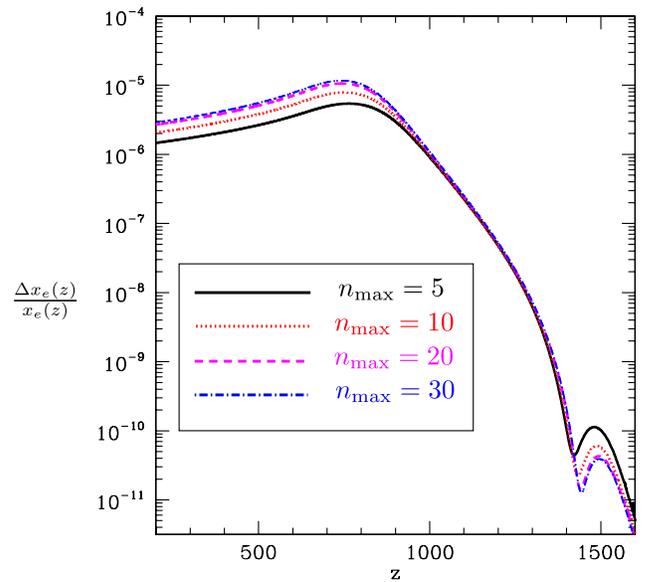}
\caption{Fractional difference between recombination histories with/without E2 quadrupole transitions included for different values of $n_{\rm max}$. The net effect is always to speed up recombination.}
\label{quad_hist}
\end{figure}

The results are shown in Fig. \ref{quad_hist}. The maximum effect of E2 transitions occurs at $z\sim 800$ with a fractional enhancement of $\Delta x_{e}/x_{e}\simeq 10^{-5}$, and the calculation seems well converged by $n_{\rm max}=30$. Corrections due to higher excited states would be a correction to a correction, and so we ignore them. Although the correction from E2 transitions is small, it extends over a broad epoch at late times after reaching its maximum. To determine if this could affect CMB anisotropies in an observable way, we modify and run CMBFast \cite{cmbfast} using recombination histories computed with/without E2 transitions. We incorporated \textsc{RecSparse} recombination histories including E2 transitions into CMBFast by applying the same method employed in Sec. \ref{resultsn}.

Running the recombination histories including E2 quadrupole transitions through {\sc CMBFast} gives a maximum change $\Delta
C_{\ell}/C_{\ell}\sim 3\times 10^{-6}$ in both temperature and polarization, negligible compared to cosmic variance.  Thus E2 transitions in hydrogen are negligible for CMB applications.

\section{Conclusions}
\label{concl}
We have developed a new recombination code, \textsc{RecSparse}, optimized for tracking the populations of many energy shells of the hydrogen atom while resolving angular momentum sublevels. The code runs more quickly than would be anticipated using simple scaling arguments, which would yield the the scaling $t_{\rm comp}\propto n_{\rm max}^{6}$. Using \textsc{RecSparse}, we find empirically that for the range of $n_{\rm max}$ values used, computation time scales as $t_{\rm comp}\propto n_{\rm max}^{\alpha}$, where $2<\alpha<3$. With this code, we have computed cosmological hydrogen recombination histories for a series of $n_{\rm max}$ values going as high as $n_{\rm max}=250$ and explored the highly nonequilibrium state of the resulting atomic hydrogen.

The resulting correction $\Delta x_{e}(z)$ satisfies $\Delta x_{e}(z)/x_{e}(z)<0.01$ for $z>200$ when $n_{\rm max}=250$ and converges with $\Delta x_{e}(z)/x_{e}(z)\propto n_{\rm max}^{-1.9}$. The correction to the $C_{\ell}$'s becomes of order the cosmic variance when $n_{\rm max}=250$. In light of realistic error estimates for \textit{Planck}, the resulting CMB anisotropy spectra $C_{\ell}^{XX}$ are converged to $0.5\sigma$ at Fisher-matrix level for $n_{\rm max}=128$ in the purely radiative case, assuming error extrapolations may be trusted. 

To definitively answer the question of absolute convergence, collisions must be included to speed the approach to Saha equilibrium at high $n$, allowing a conclusive treatment of states beyond the truncation limit, with $n>n_{\rm max}$. Future work should also properly account for the overlap of the Lyman resonance line series at high $n$. It will also be interesting to determine if there is coherent stimulated emission between excited states, given its relevance for the detectability of faint CMB spectral distortions from the epoch of recombination. Finally, the sparse-matrix methods applied here or similar techniques could be profitably applied in the development of fast recombination codes for CMB data analysis, even at early times in recombination, when only lower values of $n_{\rm max}$ are relevant.
\begin{acknowledgments}
The authors acknowledge useful conversations with Y. Ali-Ha\"{i}moud, N. Bode, A. Pullen, T. L. Smith, J. Chluba, J. A. Rubi\~{n}o-Mart\'{i}n, and the participants of the July 2009 Paris Workshop on Cosmological Recombination. D.G. is supported by the Dan David Foundation and the Gordon and Betty Moore Foundation. C.H. is supported by DoE DE-FG03-92-ER40701, the National Science Foundation under Contract No. AST-0807337, and the Alfred P. Sloan Foundation.

\end{acknowledgments}
  \renewcommand{\theequation}{A\arabic{equation}}
  \setcounter{equation}{0} 
\appendix
\section*{Appendix: WKB approximation for radial dipole integrals}
The development of laser spectroscopy of high-$n$ states in hydrogen and other atoms, along with the study of nonlinear and multiphoton ionization, required the computation of dipole radial matrix elements for high and even fractional quantum numbers in a Coulomb or perturbed Coulomb potential \cite{kaulakys_wkb}. Until adequate algorithms for these computations were ultimately developed, the Wentzel, Kramers, Brillouin, and Jeffreys (WKBJ) semiclassical approximation (quite accurate for $n\gg1$) \cite{hirata_helium_b,heim_bb_wkb_rates} proved a useful tool for estimating $\lsup{X_{n^{\prime},n}^{l^{\prime},l}}{\left(n\right)}$. At high $n$, radial wave functions in the Coulomb potential have a large number of nodes and thus a short wavelength $\lambda$. For the WKB approximation to be valid, it is necessary that $|d\lambda/dx|\ll2\pi$. Because of the large number of nodes in the Coulomb wave functions at high $n$, the WKB approximation is ideally suited to estimating matrix elements for transitions between high $n$.

\label{appendix_wkb}
In the classically allowed region, the nonrelativistic WKB radial wave function for a hydrogen atom is 
\begin{align}
&xR_{nl}(x)=\left(\frac{2}{\pi n^{3}k(x) }\right)^{1/2}\cos{\left[\int_{x_{1}}^{x} k_{nl}(x)dx-\frac{\pi}{4}\right]}\end{align} with
\begin{align}k_{nl}(x)=\left[\frac{1}{n^{2}}+\frac{2}{x}-\frac{l\left(l+1\right)}{x^{2}}\right]^{1/2},\label{wkb_bb_wfb}
\end{align}
where the inner classical turning point $x_{1}$ is a solution of the equation $k_{nl}(x)=0$. Substituting Eq.~(\ref{wkb_bb_wfb}) into Eq.~(\ref{mel_bb}) for the dipole matrix element, and making several additional approximations, the following expression is obtained if $|n^{\prime}-n|\ll n,n^{\prime}$ and $n,n^{\prime}\gg l$ \cite{heim_bb_wkb_rates}:
\begin{eqnarray}
\lsup{X_{n^{\prime},n}^{l^{\prime},l}}{\left(1\right)}=\frac{n_{c}^{2}}{2s}\left[\left(1+\Delta l \frac{l_{c}}{n_{c}}\right)J_{s+1}\left(\epsilon s\right)\right.\nonumber \\
-\left.\left(1-\Delta l \frac{l_{c}}{n_{c}}\right)J_{s-1}\left(\epsilon s \right)\right]\label{bb_low_wkb}
.\end{eqnarray} with $s=n-n^{\prime}$, $\Delta l=l^{\prime}-l$, $l_{c}=\left(l+l^{\prime}+1\right)/2$, $n_{c}=2nn^{\prime}/\left(n+n^{\prime}\right)$, and $\epsilon^{2}=1-\left(l_{c}^{2}/n_{c}^{2}\right).$ Here $\epsilon$ is the eccentricity of a Keplerian orbit with the quantum numbers $n_{c}$ and $l_{c}$, and $J_{s}\left(x\right)$ is a Bessel function of the first kind. These estimates agree with matrix elements computed using Eq.~(\ref{hga}) to a precision of $5\%$--$50\%$; the agreement worsens as $|n^{\prime}-n|\to n,n^{\prime}$. 

If $l\ll n^{\prime},n$ and $s\sim n,n^{\prime}$, then \cite{delone_bf_wkb_rates_a}
\begin{align}
&\lsup{X_{n^{\prime},n}^{l\pm 1,l}}{\left(1\right)}=2\frac{l^{2}}{\pi\sqrt{3}}\left(nn^{\prime}\right)^{-3/2}y^{-1}\nonumber \\
&\times \left\{K_{2/3}\left(\frac{l^{3}y}{6}\right)\mp K_{1/3}\left(\frac{l^{3}y}{6}\right)\right\}, \label{wkbfar} \end{align} with
$y=\left|n^{-2}-n^{\prime2} \right|$. Here $K_{s}(x)$ is a modified Bessel function of the second kind. These estimates agree with matrix elements computed using Eq.~(\ref{hga}) to a precision of $1\%$--$20\%$; the agreement worsens as $s$ shrinks, at which point Eq.~(\ref{bb_low_wkb}) becomes more accurate. 

A WKB estimate of bound-free matrix elements is obtained by making the substitution $n^{\prime}\to i/\kappa$ in Eq.~(\ref{wkbfar}) \cite{delone_bf_wkb_rates_a}. The resulting estimate is reasonable if $l\ll n,\kappa^{-1}$ and agrees with matrix elements computed using Eq. (\ref{befmel}) to a precision of $50\%$.
This analysis confirms that the high $n$ and $l$ values under consideration do not afflict our evaluation of Eqs.~(\ref{hga}) or (\ref{befmel}) with any instability that would throw computed rates off by orders of magnitude.


\begin{thebibliography}{100}
\expandafter\ifx\csname natexlab\endcsname\relax\def\natexlab#1{#1}\fi
\expandafter\ifx\csname bibnamefont\endcsname\relax
  \def\bibnamefont#1{#1}\fi
\expandafter\ifx\csname bibfnamefont\endcsname\relax
  \def\bibfnamefont#1{#1}\fi
\expandafter\ifx\csname citenamefont\endcsname\relax
  \def\citenamefont#1{#1}\fi
\expandafter\ifx\csname url\endcsname\relax
  \def\url#1{\texttt{#1}}\fi
\expandafter\ifx\csname urlprefix\endcsname\relax\def\urlprefix{URL }\fi
\providecommand{\bibinfo}[2]{#2}
\providecommand{\eprint}[2][]{\url{#2}}

\bibitem[{\citenamefont{{Hinshaw} et~al.}(2009)}]{wmap_5}
\bibinfo{author}{\bibfnamefont{G.}~\bibnamefont{{Hinshaw}}}
  \bibnamefont{et~al.}, \bibinfo{journal}{\apjs}
  \textbf{\bibinfo{volume}{180}}, \bibinfo{pages}{225} (\bibinfo{year}{2009}),
  \eprint{arXiv:0803.0732}.

\bibitem[{\citenamefont{{Steigman}}(2007)}]{steigmanreview}
\bibinfo{author}{\bibfnamefont{G.}~\bibnamefont{{Steigman}}},
  \bibinfo{journal}{Annu. Rev. Nucl. Part. Sci.} \textbf{\bibinfo{volume}{57}},
  \bibinfo{pages}{463} (\bibinfo{year}{2007}), \eprint{arXiv:0712.1100}.

\bibitem[{\citenamefont{{Perlmutter} et~al.}(1999)}]{perlmutter_supernovae}
\bibinfo{author}{\bibfnamefont{S.}~\bibnamefont{{Perlmutter}}}
  \bibnamefont{et~al.}, \bibinfo{journal}{\apj} \textbf{\bibinfo{volume}{517}},
  \bibinfo{pages}{565} (\bibinfo{year}{1999}), \eprint{arXiv:astro-ph/9812133}.

\bibitem[{\citenamefont{{Riess} et~al.}(1998)}]{reiss_supernovae}
\bibinfo{author}{\bibfnamefont{A.~G.} \bibnamefont{{Riess}}}
  \bibnamefont{et~al.}, \bibinfo{journal}{\aj} \textbf{\bibinfo{volume}{116}},
  \bibinfo{pages}{1009} (\bibinfo{year}{1998}),
  \eprint{arXiv:astro-ph/9805201}.

\bibitem[{\citenamefont{{Tegmark} et~al.}(2004)}]{sdss_pspec}
\bibinfo{author}{\bibfnamefont{M.}~\bibnamefont{{Tegmark}}}
  \bibnamefont{et~al.}, \bibinfo{journal}{\apj} \textbf{\bibinfo{volume}{606}},
  \bibinfo{pages}{702} (\bibinfo{year}{2004}), \eprint{arXiv:astro-ph/0310725}.

\bibitem[{\citenamefont{{Padmanabhan} et~al.}(2007)}]{sdss_lrg}
\bibinfo{author}{\bibfnamefont{N.}~\bibnamefont{{Padmanabhan}}}
  \bibnamefont{et~al.}, \bibinfo{journal}{\mnras}
  \textbf{\bibinfo{volume}{378}}, \bibinfo{pages}{852} (\bibinfo{year}{2007}),
  \eprint{arXiv:astro-ph/0605302}.

\bibitem[{\citenamefont{{Cole} et~al.}(2005)}]{bao_discovery_cole}
\bibinfo{author}{\bibfnamefont{S.}~\bibnamefont{{Cole}}} \bibnamefont{et~al.},
  \bibinfo{journal}{\mnras} \textbf{\bibinfo{volume}{362}},
  \bibinfo{pages}{505} (\bibinfo{year}{2005}), \eprint{arXiv:astro-ph/0501174}.

\bibitem[{\citenamefont{{Rozo} et~al.}(2010)}]{cluster_cosmo}
\bibinfo{author}{\bibfnamefont{E.}~\bibnamefont{{Rozo}}} \bibnamefont{et~al.},
  \bibinfo{journal}{\apj} \textbf{\bibinfo{volume}{708}}, \bibinfo{pages}{645}
  (\bibinfo{year}{2010}), \eprint{0902.3702}.

\bibitem[{\citenamefont{{Caldwell} et~al.}(1998)\citenamefont{{Caldwell},
  {Dave}, and {Steinhardt}}}]{caldwell_darkenergy}
\bibinfo{author}{\bibfnamefont{R.~R.} \bibnamefont{{Caldwell}}},
  \bibinfo{author}{\bibfnamefont{R.}~\bibnamefont{{Dave}}}, \bibnamefont{and}
  \bibinfo{author}{\bibfnamefont{P.~J.} \bibnamefont{{Steinhardt}}},
  \bibinfo{journal}{\prl} \textbf{\bibinfo{volume}{80}}, \bibinfo{pages}{1582}
  (\bibinfo{year}{1998}), \eprint{arXiv:astro-ph/9708069}.

\bibitem[{\citenamefont{{Carroll} et~al.}(2004)\citenamefont{{Carroll},
  {Duvvuri}, {Trodden}, and {Turner}}}]{carroll_1r}
\bibinfo{author}{\bibfnamefont{S.~M.} \bibnamefont{{Carroll}}},
  \bibinfo{author}{\bibfnamefont{V.}~\bibnamefont{{Duvvuri}}},
  \bibinfo{author}{\bibfnamefont{M.}~\bibnamefont{{Trodden}}},
  \bibnamefont{and} \bibinfo{author}{\bibfnamefont{M.~S.}
  \bibnamefont{{Turner}}}, \bibinfo{journal}{\prd}
  \textbf{\bibinfo{volume}{70}}, \bibinfo{pages}{043528}
  (\bibinfo{year}{2004}), \eprint{arXiv:astro-ph/0306438}.

\bibitem[{\citenamefont{{Ichikawa} et~al.}(2005)\citenamefont{{Ichikawa},
  {Fukugita}, and {Kawasaki}}}]{neutrino}
\bibinfo{author}{\bibfnamefont{K.}~\bibnamefont{{Ichikawa}}},
  \bibinfo{author}{\bibfnamefont{M.}~\bibnamefont{{Fukugita}}},
  \bibnamefont{and}
  \bibinfo{author}{\bibfnamefont{M.}~\bibnamefont{{Kawasaki}}},
  \bibinfo{journal}{\prd} \textbf{\bibinfo{volume}{71}},
  \bibinfo{pages}{043001} (\bibinfo{year}{2005}),
  \eprint{arXiv:astro-ph/0409768}.

\bibitem[{\citenamefont{{Dodelson} et~al.}(1996)\citenamefont{{Dodelson},
  {Gates}, and {Stebbins}}}]{neutrinob}
\bibinfo{author}{\bibfnamefont{S.}~\bibnamefont{{Dodelson}}},
  \bibinfo{author}{\bibfnamefont{E.}~\bibnamefont{{Gates}}}, \bibnamefont{and}
  \bibinfo{author}{\bibfnamefont{A.}~\bibnamefont{{Stebbins}}},
  \bibinfo{journal}{\apj} \textbf{\bibinfo{volume}{467}}, \bibinfo{pages}{10}
  (\bibinfo{year}{1996}), \eprint{arXiv:astro-ph/9509147}.

\bibitem[{\citenamefont{{Ma} and {Bertschinger}}(1995)}]{mabert}
\bibinfo{author}{\bibfnamefont{C.-P.} \bibnamefont{{Ma}}} \bibnamefont{and}
  \bibinfo{author}{\bibfnamefont{E.}~\bibnamefont{{Bertschinger}}},
  \bibinfo{journal}{\apj} \textbf{\bibinfo{volume}{455}}, \bibinfo{pages}{7}
  (\bibinfo{year}{1995}), \eprint{arXiv:astro-ph/9506072}.

\bibitem[{\citenamefont{{Netterfield} et~al.}(2002)}]{boomerang_b}
\bibinfo{author}{\bibfnamefont{C.~B.} \bibnamefont{{Netterfield}}}
  \bibnamefont{et~al.}, \bibinfo{journal}{\apj} \textbf{\bibinfo{volume}{571}},
  \bibinfo{pages}{604} (\bibinfo{year}{2002}), \eprint{arXiv:astro-ph/0104460}.

\bibitem[{\citenamefont{{Pearson} et~al.}(2003)}]{cbi}
\bibinfo{author}{\bibfnamefont{T.~J.} \bibnamefont{{Pearson}}}
  \bibnamefont{et~al.}, \bibinfo{journal}{\apj} \textbf{\bibinfo{volume}{591}},
  \bibinfo{pages}{556} (\bibinfo{year}{2003}), \eprint{arXiv:astro-ph/0205388}.

\bibitem[{\citenamefont{{Kuo} et~al.}(2004)}]{acbar_a}
\bibinfo{author}{\bibfnamefont{C.~L.} \bibnamefont{{Kuo}}}
  \bibnamefont{et~al.}, \bibinfo{journal}{\apj} \textbf{\bibinfo{volume}{600}},
  \bibinfo{pages}{32} (\bibinfo{year}{2004}), \eprint{arXiv:astro-ph/0212289}.

\bibitem[{\citenamefont{Reichardt et~al.}(2009)}]{acbar_c}
\bibinfo{author}{\bibfnamefont{C.~L.} \bibnamefont{Reichardt}}
  \bibnamefont{et~al.}, \bibinfo{journal}{\apj} \textbf{\bibinfo{volume}{694}},
  \bibinfo{pages}{1200} (\bibinfo{year}{2009}), \eprint{arXiv:0801.1491}.

\bibitem[{\citenamefont{{Kovac} et~al.}(2002)}]{dasi_a}
\bibinfo{author}{\bibfnamefont{J.~M.} \bibnamefont{{Kovac}}}
  \bibnamefont{et~al.}, \bibinfo{journal}{\natu}
  \textbf{\bibinfo{volume}{420}}, \bibinfo{pages}{772} (\bibinfo{year}{2002}),
  \eprint{arXiv:astro-ph/0209478}.

\bibitem[{\citenamefont{{Chiang} et~al.}(2010)}]{bicep}
\bibinfo{author}{\bibfnamefont{H.~C.} \bibnamefont{{Chiang}}}
  \bibnamefont{et~al.}, \bibinfo{journal}{\apj} \textbf{\bibinfo{volume}{711}},
  \bibinfo{pages}{1123} (\bibinfo{year}{2010}), \eprint{0906.1181}.

\bibitem[{\citenamefont{Seljak and Zaldarriaga}(1997)}]{seljak_pol}
\bibinfo{author}{\bibfnamefont{U.}~\bibnamefont{Seljak}} \bibnamefont{and}
  \bibinfo{author}{\bibfnamefont{M.}~\bibnamefont{Zaldarriaga}},
  \bibinfo{journal}{\prl} \textbf{\bibinfo{volume}{78}}, \bibinfo{pages}{2054}
  (\bibinfo{year}{1997}), \eprint{astro-ph/9609169}.

\bibitem[{\citenamefont{Kamionkowski et~al.}(1997)\citenamefont{Kamionkowski,
  Kosowsky, and Stebbins}}]{kamion_pol}
\bibinfo{author}{\bibfnamefont{M.}~\bibnamefont{Kamionkowski}},
  \bibinfo{author}{\bibfnamefont{A.}~\bibnamefont{Kosowsky}}, \bibnamefont{and}
  \bibinfo{author}{\bibfnamefont{A.}~\bibnamefont{Stebbins}},
  \bibinfo{journal}{\prd} \textbf{\bibinfo{volume}{55}}, \bibinfo{pages}{7368}
  (\bibinfo{year}{1997}), \eprint{astro-ph/9611125}.

\bibitem[{\citenamefont{{The Planck Collaboration}}(2006)}]{planck}
\bibinfo{author}{\bibnamefont{{The Planck Collaboration}}}
  (\bibinfo{year}{2006}), \eprint{arXiv:astro-ph/0604069}.

\bibitem[{\citenamefont{{Albrecht} et~al.}(2009)}]{detf_b}
\bibinfo{author}{\bibfnamefont{A.}~\bibnamefont{{Albrecht}}}
  \bibnamefont{et~al.} (\bibinfo{year}{2009}), \eprint{0901.0721}.

\bibitem[{\citenamefont{{Eisenstein} et~al.}(2005)}]{bao_discovery_eisenstein}
\bibinfo{author}{\bibfnamefont{D.~J.} \bibnamefont{{Eisenstein}}}
  \bibnamefont{et~al.}, \bibinfo{journal}{\apj} \textbf{\bibinfo{volume}{633}},
  \bibinfo{pages}{560} (\bibinfo{year}{2005}), \eprint{arXiv:astro-ph/0501171}.

\bibitem[{\citenamefont{{Ruhl} et~al.}(2004)}]{spt}
\bibinfo{author}{\bibfnamefont{J.}~\bibnamefont{{Ruhl}}} \bibnamefont{et~al.},
  in \emph{\bibinfo{booktitle}{Society of Photo-Optical Instrumentation
  Engineers (SPIE) Conference Series}}, edited by
  \bibinfo{editor}{\bibfnamefont{C.~M.} \bibnamefont{{Bradford}}}
  \bibnamefont{et~al.} (\bibinfo{year}{2004}), vol. \bibinfo{volume}{5498} of
  \emph{\bibinfo{series}{Society of Photo-Optical Instrumentation Engineers
  (SPIE) Conference Series}}, pp. \bibinfo{pages}{11--29}.

\bibitem[{\citenamefont{{Kosowsky}}(2003)}]{act}
\bibinfo{author}{\bibfnamefont{A.}~\bibnamefont{{Kosowsky}}},
  \bibinfo{journal}{New Astron. Rev.} \textbf{\bibinfo{volume}{47}},
  \bibinfo{pages}{939} (\bibinfo{year}{2003}), \eprint{arXiv:astro-ph/0402234}.

\bibitem[{\citenamefont{Zaldarriaga et~al.}(2008)}]{cmbpol_c}
\bibinfo{author}{\bibfnamefont{M.}~\bibnamefont{Zaldarriaga}}
  \bibnamefont{et~al.} (\bibinfo{year}{2008}), \eprint{arXiv:0811.3918}.

\bibitem[{\citenamefont{Baumann et~al.}(2009)}]{cmbpol_d}
\bibinfo{author}{\bibfnamefont{D.}~\bibnamefont{Baumann}} \bibnamefont{et~al.}
  (\bibinfo{collaboration}{CMBPol Study Team}), \bibinfo{journal}{AIP Conf.
  Proc.} \textbf{\bibinfo{volume}{1141}}, \bibinfo{pages}{10}
  (\bibinfo{year}{2009}), \eprint{arXiv:0811.3919}.

\bibitem[{\citenamefont{{Bond} and {Efstathiou}}(1987)}]{bond_acoustic}
\bibinfo{author}{\bibfnamefont{J.~R.} \bibnamefont{{Bond}}} \bibnamefont{and}
  \bibinfo{author}{\bibfnamefont{G.}~\bibnamefont{{Efstathiou}}},
  \bibinfo{journal}{\mnras} \textbf{\bibinfo{volume}{226}},
  \bibinfo{pages}{655} (\bibinfo{year}{1987}).

\bibitem[{\citenamefont{{Peebles} and {Yu}}(1970)}]{peebles_acoustic}
\bibinfo{author}{\bibfnamefont{P.~J.~E.} \bibnamefont{{Peebles}}}
  \bibnamefont{and} \bibinfo{author}{\bibfnamefont{J.~T.} \bibnamefont{{Yu}}},
  \bibinfo{journal}{\apj} \textbf{\bibinfo{volume}{162}}, \bibinfo{pages}{815}
  (\bibinfo{year}{1970}).

\bibitem[{\citenamefont{{Silk}}(1968)}]{silk_damp}
\bibinfo{author}{\bibfnamefont{J.}~\bibnamefont{{Silk}}},
  \bibinfo{journal}{\apj} \textbf{\bibinfo{volume}{151}}, \bibinfo{pages}{459}
  (\bibinfo{year}{1968}).

\bibitem[{\citenamefont{{Hu}}(1997)}]{hu_sugiyama_silk_acoustic}
\bibinfo{author}{\bibfnamefont{W.}~\bibnamefont{{Hu}}}, \bibinfo{journal}{\nat}
  \textbf{\bibinfo{volume}{386}}, \bibinfo{pages}{37} (\bibinfo{year}{1997}),
  \eprint{arXiv:astro-ph/9504057}.

\bibitem[{\citenamefont{{Bond} and {Efstathiou}}(1984)}]{bond_pol}
\bibinfo{author}{\bibfnamefont{J.~R.} \bibnamefont{{Bond}}} \bibnamefont{and}
  \bibinfo{author}{\bibfnamefont{G.}~\bibnamefont{{Efstathiou}}},
  \bibinfo{journal}{\apjl} \textbf{\bibinfo{volume}{285}}, \bibinfo{pages}{L45}
  (\bibinfo{year}{1984}).

\bibitem[{\citenamefont{{Polnarev}}(1985)}]{polnarev_pol_a}
\bibinfo{author}{\bibfnamefont{A.~G.} \bibnamefont{{Polnarev}}},
  \bibinfo{journal}{\sa} \textbf{\bibinfo{volume}{29}}, \bibinfo{pages}{607}
  (\bibinfo{year}{1985}).

\bibitem[{\citenamefont{{Scott}}(1999)}]{scottsmear}
\bibinfo{author}{\bibfnamefont{D.}~\bibnamefont{{Scott}}}, in
  \emph{\bibinfo{booktitle}{Evolution of Large Scale Structure : From
  Recombination to Garching}}, edited by \bibinfo{editor}{\bibfnamefont{A.~J.}
  \bibnamefont{{Banday}}}, \bibinfo{editor}{\bibfnamefont{R.~K.}
  \bibnamefont{{Sheth}}}, \bibnamefont{and}
  \bibinfo{editor}{\bibfnamefont{L.~N.} \bibnamefont{{da Costa}}}
  (\bibinfo{year}{1999}), pp. \bibinfo{pages}{30--+}.

\bibitem[{\citenamefont{{Seager} et~al.}(1999)\citenamefont{{Seager},
  {Sasselov}, and {Scott}}}]{seager_recfast}
\bibinfo{author}{\bibfnamefont{S.}~\bibnamefont{{Seager}}},
  \bibinfo{author}{\bibfnamefont{D.~D.} \bibnamefont{{Sasselov}}},
  \bibnamefont{and} \bibinfo{author}{\bibfnamefont{D.}~\bibnamefont{{Scott}}},
  \bibinfo{journal}{\apjl} \textbf{\bibinfo{volume}{523}}, \bibinfo{pages}{L1}
  (\bibinfo{year}{1999}), \eprint{arXiv:astro-ph/9909275}.

\bibitem[{\citenamefont{{Lewis} et~al.}(2006)\citenamefont{{Lewis}, {Weller},
  and {Battye}}}]{lewis_rec_cmb_param}
\bibinfo{author}{\bibfnamefont{A.}~\bibnamefont{{Lewis}}},
  \bibinfo{author}{\bibfnamefont{J.}~\bibnamefont{{Weller}}}, \bibnamefont{and}
  \bibinfo{author}{\bibfnamefont{R.}~\bibnamefont{{Battye}}},
  \bibinfo{journal}{\mnras} \textbf{\bibinfo{volume}{373}},
  \bibinfo{pages}{561} (\bibinfo{year}{2006}), \eprint{arXiv:astro-ph/0606552}.

\bibitem[{\citenamefont{{Wong} et~al.}(2008)\citenamefont{{Wong}, {Moss}, and
  {Scott}}}]{wong_better_rec_b}
\bibinfo{author}{\bibfnamefont{W.~Y.} \bibnamefont{{Wong}}},
  \bibinfo{author}{\bibfnamefont{A.}~\bibnamefont{{Moss}}}, \bibnamefont{and}
  \bibinfo{author}{\bibfnamefont{D.}~\bibnamefont{{Scott}}},
  \bibinfo{journal}{\mnras} \textbf{\bibinfo{volume}{386}},
  \bibinfo{pages}{1023} (\bibinfo{year}{2008}), \eprint{arXiv:0711.1357}.

\bibitem[{\citenamefont{{Wong} and {Scott}}(2006)}]{improving_wong}
\bibinfo{author}{\bibfnamefont{W.~Y.} \bibnamefont{{Wong}}} \bibnamefont{and}
  \bibinfo{author}{\bibfnamefont{D.}~\bibnamefont{{Scott}}}, in
  \emph{\bibinfo{booktitle}{Bull. Am. Astron. Soc.}} (\bibinfo{year}{2006}),
  vol.~\bibinfo{volume}{38} of \emph{\bibinfo{series}{Bulletin of the American
  Astronomical Society}}, pp. \bibinfo{pages}{1210--+}.

\bibitem[{\citenamefont{{Peebles}}(1968)}]{peebles_rec}
\bibinfo{author}{\bibfnamefont{P.~J.~E.} \bibnamefont{{Peebles}}},
  \bibinfo{journal}{\apj} \textbf{\bibinfo{volume}{153}}, \bibinfo{pages}{1}
  (\bibinfo{year}{1968}).

\bibitem[{\citenamefont{{Zel'dovich} et~al.}(1968)\citenamefont{{Zel'dovich},
  {Kurt}, and {Sunyaev}}}]{zeldovich_rec}
\bibinfo{author}{\bibfnamefont{Y.~B.} \bibnamefont{{Zel'dovich}}},
  \bibinfo{author}{\bibfnamefont{V.~G.} \bibnamefont{{Kurt}}},
  \bibnamefont{and} \bibinfo{author}{\bibfnamefont{R.~A.}
  \bibnamefont{{Sunyaev}}}, \bibinfo{journal}{Sov.~Phys.~JETP}
  \textbf{\bibinfo{volume}{28}}, \bibinfo{pages}{146} (\bibinfo{year}{1968}).

\bibitem[{\citenamefont{{Rybicki} and
  {dell'Antonio}}(1993)}]{rybicki_specdist_a}
\bibinfo{author}{\bibfnamefont{G.~B.} \bibnamefont{{Rybicki}}}
  \bibnamefont{and} \bibinfo{author}{\bibfnamefont{I.~P.}
  \bibnamefont{{dell'Antonio}}}, in \emph{\bibinfo{booktitle}{Observational
  Cosmology}}, edited by \bibinfo{editor}{\bibfnamefont{G.~L.}
  \bibnamefont{{Chincarini}}},
  \bibinfo{editor}{\bibfnamefont{A.}~\bibnamefont{{Iovino}}},
  \bibinfo{editor}{\bibfnamefont{T.}~\bibnamefont{{Maccacaro}}},
  \bibnamefont{and}
  \bibinfo{editor}{\bibfnamefont{D.}~\bibnamefont{{Maccagni}}}
  (\bibinfo{year}{1993}), vol.~\bibinfo{volume}{51} of
  \emph{\bibinfo{series}{Astronomical Society of the Pacific Conference
  Series}}, pp. \bibinfo{pages}{548--+}.

\bibitem[{\citenamefont{{Dubrovich}}(1975)}]{dubrovich_a}
\bibinfo{author}{\bibfnamefont{V.~K.} \bibnamefont{{Dubrovich}}},
  \bibinfo{journal}{\sa} \textbf{\bibinfo{volume}{1}}, \bibinfo{pages}{3}
  (\bibinfo{year}{1975}).

\bibitem[{\citenamefont{{Seager} et~al.}(2000)\citenamefont{{Seager},
  {Sasselov}, and {Scott}}}]{seager_phys_recfast}
\bibinfo{author}{\bibfnamefont{S.}~\bibnamefont{{Seager}}},
  \bibinfo{author}{\bibfnamefont{D.~D.} \bibnamefont{{Sasselov}}},
  \bibnamefont{and} \bibinfo{author}{\bibfnamefont{D.}~\bibnamefont{{Scott}}},
  \bibinfo{journal}{\apjs} \textbf{\bibinfo{volume}{128}}, \bibinfo{pages}{407}
  (\bibinfo{year}{2000}), \eprint{arXiv:astro-ph/9912182}.

\bibitem[{\citenamefont{Weymann}(1965)}]{weymann_tmtr}
\bibinfo{author}{\bibfnamefont{R.}~\bibnamefont{Weymann}},
  \bibinfo{journal}{Phys. Fluids} \textbf{\bibinfo{volume}{8}},
  \bibinfo{pages}{2112} (\bibinfo{year}{1965}),
  \urlprefix\url{http://link.aip.org/link/?PFL/8/2112/1}.

\bibitem[{\citenamefont{Sunyaev and Zel'dovich}(1970)}]{sunyaev_tmtr}
\bibinfo{author}{\bibfnamefont{R.~A.} \bibnamefont{Sunyaev}} \bibnamefont{and}
  \bibinfo{author}{\bibfnamefont{Y.~B.} \bibnamefont{Zel'dovich}},
  \bibinfo{journal}{Astrophys. Space Sci.} \textbf{\bibinfo{volume}{7}},
  \bibinfo{pages}{20} (\bibinfo{year}{1970}).

\bibitem[{\citenamefont{{Sobolev}}(1957)}]{sobolev_sobolev_a}
\bibinfo{author}{\bibfnamefont{V.~V.} \bibnamefont{{Sobolev}}},
  \bibinfo{journal}{\sa} \textbf{\bibinfo{volume}{1}}, \bibinfo{pages}{332}
  (\bibinfo{year}{1957}).

\bibitem[{\citenamefont{{Kholupenko} and
  {Ivanchik}}(2006)}]{kholupenko_twophoton}
\bibinfo{author}{\bibfnamefont{E.~E.} \bibnamefont{{Kholupenko}}}
  \bibnamefont{and} \bibinfo{author}{\bibfnamefont{A.~V.}
  \bibnamefont{{Ivanchik}}}, \bibinfo{journal}{Astron. Lett.}
  \textbf{\bibinfo{volume}{32}}, \bibinfo{pages}{795} (\bibinfo{year}{2006}),
  \eprint{arXiv:astro-ph/0611395}.

\bibitem[{\citenamefont{{Chluba} and {Sunyaev}}(2006)}]{chluba_twophoton_a}
\bibinfo{author}{\bibfnamefont{J.}~\bibnamefont{{Chluba}}} \bibnamefont{and}
  \bibinfo{author}{\bibfnamefont{R.~A.} \bibnamefont{{Sunyaev}}},
  \bibinfo{journal}{\aap} \textbf{\bibinfo{volume}{446}}, \bibinfo{pages}{39}
  (\bibinfo{year}{2006}), \eprint{arXiv:astro-ph/0508144}.

\bibitem[{\citenamefont{{Chluba} and {Sunyaev}}(2008)}]{chluba_twophoton_b}
\bibinfo{author}{\bibfnamefont{J.}~\bibnamefont{{Chluba}}} \bibnamefont{and}
  \bibinfo{author}{\bibfnamefont{R.~A.} \bibnamefont{{Sunyaev}}},
  \bibinfo{journal}{\aap} \textbf{\bibinfo{volume}{480}}, \bibinfo{pages}{629}
  (\bibinfo{year}{2008}), \eprint{arXiv:0705.3033}.

\bibitem[{\citenamefont{Hirata}(2008)}]{hirata_twophoton}
\bibinfo{author}{\bibfnamefont{C.~M.} \bibnamefont{Hirata}},
  \bibinfo{journal}{\prd} \textbf{\bibinfo{volume}{78}},
  \bibinfo{pages}{023001} (\bibinfo{year}{2008}), \eprint{arXiv:0803.0808}.

\bibitem[{\citenamefont{{Chluba} and
  {Sunyaev}}(2009{\natexlab{a}})}]{chluba_lymana_b}
\bibinfo{author}{\bibfnamefont{J.}~\bibnamefont{{Chluba}}} \bibnamefont{and}
  \bibinfo{author}{\bibfnamefont{R.~A.} \bibnamefont{{Sunyaev}}}
  (\bibinfo{year}{2009}{\natexlab{a}}), \eprint{arXiv:0904.0460}.

\bibitem[{\citenamefont{{Wong} and {Scott}}(2007)}]{wong_scott_forbidden}
\bibinfo{author}{\bibfnamefont{W.~Y.} \bibnamefont{{Wong}}} \bibnamefont{and}
  \bibinfo{author}{\bibfnamefont{D.}~\bibnamefont{{Scott}}},
  \bibinfo{journal}{\mnras} \textbf{\bibinfo{volume}{375}},
  \bibinfo{pages}{1441} (\bibinfo{year}{2007}),
  \eprint{arXiv:astro-ph/0610691}.

\bibitem[{\citenamefont{Hirata and Switzer}(2008)}]{hirata_helium_b}
\bibinfo{author}{\bibfnamefont{C.~M.} \bibnamefont{Hirata}} \bibnamefont{and}
  \bibinfo{author}{\bibfnamefont{E.~R.} \bibnamefont{Switzer}},
  \bibinfo{journal}{\prd} \textbf{\bibinfo{volume}{77}},
  \bibinfo{pages}{083007} (\bibinfo{year}{2008}), \eprint{astro-ph/0702144}.

\bibitem[{\citenamefont{{Dubrovich} and {Grachev}}(2005)}]{helium_forbidden}
\bibinfo{author}{\bibfnamefont{V.~K.} \bibnamefont{{Dubrovich}}}
  \bibnamefont{and} \bibinfo{author}{\bibfnamefont{S.~I.}
  \bibnamefont{{Grachev}}}, \bibinfo{journal}{Astron. Lett.}
  \textbf{\bibinfo{volume}{31}}, \bibinfo{pages}{359} (\bibinfo{year}{2005}),
  \eprint{{arXiv:astro-ph/0501672}}.

\bibitem[{\citenamefont{{Chluba} and {Sunyaev}}(2007)}]{chluba_lyman_feedback}
\bibinfo{author}{\bibfnamefont{J.}~\bibnamefont{{Chluba}}} \bibnamefont{and}
  \bibinfo{author}{\bibfnamefont{R.~A.} \bibnamefont{{Sunyaev}}},
  \bibinfo{journal}{\aap} \textbf{\bibinfo{volume}{475}}, \bibinfo{pages}{109}
  (\bibinfo{year}{2007}), \eprint{arXiv:astro-ph/0702531}.

\bibitem[{\citenamefont{{Kholupenko} et~al.}(2007)\citenamefont{{Kholupenko},
  {Ivanchik}, and {Varshalovich}}}]{kholupenko_helium_a}
\bibinfo{author}{\bibfnamefont{E.~E.} \bibnamefont{{Kholupenko}}},
  \bibinfo{author}{\bibfnamefont{A.~V.} \bibnamefont{{Ivanchik}}},
  \bibnamefont{and} \bibinfo{author}{\bibfnamefont{D.~A.}
  \bibnamefont{{Varshalovich}}}, \bibinfo{journal}{\mnras}
  \textbf{\bibinfo{volume}{378}}, \bibinfo{pages}{L39} (\bibinfo{year}{2007}),
  \eprint{arXiv:astro-ph/0703438}.

\bibitem[{\citenamefont{{Sunyaev} and {Chluba}}(2007)}]{sunyaev_helium_a}
\bibinfo{author}{\bibfnamefont{R.~A.} \bibnamefont{{Sunyaev}}}
  \bibnamefont{and} \bibinfo{author}{\bibfnamefont{J.}~\bibnamefont{{Chluba}}},
  \bibinfo{journal}{Nuovo Cimento B} \textbf{\bibinfo{volume}{122}},
  \bibinfo{pages}{919} (\bibinfo{year}{2007}), \eprint{arXiv:0802.0772}.

\bibitem[{\citenamefont{{Chluba} and
  {Sunyaev}}(2009{\natexlab{b}})}]{chluba_lymana_a}
\bibinfo{author}{\bibfnamefont{J.}~\bibnamefont{{Chluba}}} \bibnamefont{and}
  \bibinfo{author}{\bibfnamefont{R.~A.} \bibnamefont{{Sunyaev}}},
  \bibinfo{journal}{\aap} \textbf{\bibinfo{volume}{496}}, \bibinfo{pages}{619}
  (\bibinfo{year}{2009}{\natexlab{b}}), \eprint{arXiv:0810.1045}.

\bibitem[{\citenamefont{{Rubi{\~n}o-Mart{\'{\i}}n}
  et~al.}(2006)\citenamefont{{Rubi{\~n}o-Mart{\'{\i}}n}, {Chluba}, and
  {Sunyaev}}}]{chluba_highn_a}
\bibinfo{author}{\bibfnamefont{J.~A.}
  \bibnamefont{{Rubi{\~n}o-Mart{\'{\i}}n}}},
  \bibinfo{author}{\bibfnamefont{J.}~\bibnamefont{{Chluba}}}, \bibnamefont{and}
  \bibinfo{author}{\bibfnamefont{R.~A.} \bibnamefont{{Sunyaev}}},
  \bibinfo{journal}{\mnras} \textbf{\bibinfo{volume}{371}},
  \bibinfo{pages}{1939} (\bibinfo{year}{2006}),
  \eprint{arXiv:astro-ph/0607373}.

\bibitem[{\citenamefont{{Chluba} et~al.}(2007)\citenamefont{{Chluba},
  {Rubi{\~n}o-Mart{\'{\i}}n}, and {Sunyaev}}}]{chluba_highn_b}
\bibinfo{author}{\bibfnamefont{J.}~\bibnamefont{{Chluba}}},
  \bibinfo{author}{\bibfnamefont{J.~A.}
  \bibnamefont{{Rubi{\~n}o-Mart{\'{\i}}n}}}, \bibnamefont{and}
  \bibinfo{author}{\bibfnamefont{R.~A.} \bibnamefont{{Sunyaev}}},
  \bibinfo{journal}{\mnras} \textbf{\bibinfo{volume}{374}},
  \bibinfo{pages}{1310} (\bibinfo{year}{2007}),
  \eprint{arXiv:astro-ph/0608242}.

\bibitem[{\citenamefont{{Hirata} and {Forbes}}(2009)}]{hirata_lymana}
\bibinfo{author}{\bibfnamefont{C.~M.} \bibnamefont{{Hirata}}} \bibnamefont{and}
  \bibinfo{author}{\bibfnamefont{J.}~\bibnamefont{{Forbes}}},
  \bibinfo{journal}{\prd} \textbf{\bibinfo{volume}{80}},
  \bibinfo{pages}{023001} (\bibinfo{year}{2009}), \eprint{0903.4925}.

\bibitem[{\citenamefont{Mohr et~al.}(2008)\citenamefont{Mohr, Taylor, and
  Newell}}]{nist}
\bibinfo{author}{\bibfnamefont{P.~J.} \bibnamefont{Mohr}},
  \bibinfo{author}{\bibfnamefont{B.~N.} \bibnamefont{Taylor}},
  \bibnamefont{and} \bibinfo{author}{\bibfnamefont{D.~B.}
  \bibnamefont{Newell}}, \bibinfo{journal}{Rev. Mod. Phys.}
  \textbf{\bibinfo{volume}{80}}, \bibinfo{pages}{633} (\bibinfo{year}{2008}),
  \eprint{arXiv:0801.0028}.

\bibitem[{\citenamefont{{Goldman}}(1989)}]{goldman}
\bibinfo{author}{\bibfnamefont{S.~P.} \bibnamefont{{Goldman}}},
  \bibinfo{journal}{\pra} \textbf{\bibinfo{volume}{40}}, \bibinfo{pages}{1185}
  (\bibinfo{year}{1989}).

\bibitem[{\citenamefont{{Krolik}}(1989)}]{krolik_lymana_a}
\bibinfo{author}{\bibfnamefont{J.~H.} \bibnamefont{{Krolik}}},
  \bibinfo{journal}{\apj} \textbf{\bibinfo{volume}{338}}, \bibinfo{pages}{594}
  (\bibinfo{year}{1989}).

\bibitem[{\citenamefont{{Rybicki} and {dell'Antonio}}(1990)}]{rybicki_lymana_a}
\bibinfo{author}{\bibfnamefont{G.~B.} \bibnamefont{{Rybicki}}}
  \bibnamefont{and} \bibinfo{author}{\bibfnamefont{I.~P.}
  \bibnamefont{{dell'Antonio}}}, in \emph{\bibinfo{booktitle}{\aas}}
  (\bibinfo{year}{1990}), vol.~\bibinfo{volume}{22}, pp.
  \bibinfo{pages}{1214--+}.

\bibitem[{\citenamefont{{Grachev} and {Dubrovich}}(2008)}]{grachev_lymana}
\bibinfo{author}{\bibfnamefont{S.~I.} \bibnamefont{{Grachev}}}
  \bibnamefont{and} \bibinfo{author}{\bibfnamefont{V.~K.}
  \bibnamefont{{Dubrovich}}}, \bibinfo{journal}{Astron. Lett.}
  \textbf{\bibinfo{volume}{34}}, \bibinfo{pages}{439} (\bibinfo{year}{2008}),
  \eprint{arXiv:0801.3347}.

\bibitem[{\citenamefont{{Chluba} and
  {Sunyaev}}(2009{\natexlab{c}})}]{chluba_lymana_plusthompson}
\bibinfo{author}{\bibfnamefont{J.}~\bibnamefont{{Chluba}}} \bibnamefont{and}
  \bibinfo{author}{\bibfnamefont{R.~A.} \bibnamefont{{Sunyaev}}}
  (\bibinfo{year}{2009}{\natexlab{c}}), \eprint{arXiv:0904.2220}.

\bibitem[{\citenamefont{Switzer and
  Hirata}(2008{\natexlab{a}})}]{hirata_helium_c}
\bibinfo{author}{\bibfnamefont{E.~R.} \bibnamefont{Switzer}} \bibnamefont{and}
  \bibinfo{author}{\bibfnamefont{C.~M.} \bibnamefont{Hirata}},
  \bibinfo{journal}{\prd} \textbf{\bibinfo{volume}{77}},
  \bibinfo{pages}{083008} (\bibinfo{year}{2008}{\natexlab{a}}),
  \eprint{arXiv:astro-ph/0702145}.

\bibitem[{\citenamefont{Switzer and
  Hirata}(2008{\natexlab{b}})}]{hirata_helium_a}
\bibinfo{author}{\bibfnamefont{E.~R.} \bibnamefont{Switzer}} \bibnamefont{and}
  \bibinfo{author}{\bibfnamefont{C.~M.} \bibnamefont{Hirata}},
  \bibinfo{journal}{\prd} \textbf{\bibinfo{volume}{77}},
  \bibinfo{pages}{083006} (\bibinfo{year}{2008}{\natexlab{b}}),
  \eprint{arXiv:astro-ph/0702143}.

\bibitem[{\citenamefont{{Brocklehurst}}(1970)}]{brocklehurst_rates_a}
\bibinfo{author}{\bibfnamefont{M.}~\bibnamefont{{Brocklehurst}}},
  \bibinfo{journal}{\mnras} \textbf{\bibinfo{volume}{148}},
  \bibinfo{pages}{417} (\bibinfo{year}{1970}).

\bibitem[{\citenamefont{{Goldberg}}(1966)}]{goldberg_a}
\bibinfo{author}{\bibfnamefont{L.}~\bibnamefont{{Goldberg}}},
  \bibinfo{journal}{\apj} \textbf{\bibinfo{volume}{144}}, \bibinfo{pages}{1225}
  (\bibinfo{year}{1966}).

\bibitem[{\citenamefont{{Seaton}}(1964)}]{seaton_spec_a}
\bibinfo{author}{\bibfnamefont{M.~J.} \bibnamefont{{Seaton}}},
  \bibinfo{journal}{\mnras} \textbf{\bibinfo{volume}{127}},
  \bibinfo{pages}{177} (\bibinfo{year}{1964}).

\bibitem[{\citenamefont{{Fendt} et~al.}(2009)\citenamefont{{Fendt}, {Chluba},
  {Rubi{\~n}o-Mart{\'{\i}}n}, and {Wandelt}}}]{chluba_rico}
\bibinfo{author}{\bibfnamefont{W.~A.} \bibnamefont{{Fendt}}},
  \bibinfo{author}{\bibfnamefont{J.}~\bibnamefont{{Chluba}}},
  \bibinfo{author}{\bibfnamefont{J.~A.}
  \bibnamefont{{Rubi{\~n}o-Mart{\'{\i}}n}}}, \bibnamefont{and}
  \bibinfo{author}{\bibfnamefont{B.~D.} \bibnamefont{{Wandelt}}},
  \bibinfo{journal}{\apjs} \textbf{\bibinfo{volume}{181}}, \bibinfo{pages}{627}
  (\bibinfo{year}{2009}), \eprint{arXiv:0807.2577}.

\bibitem[{\citenamefont{{Hoang-Binh}}(1990)}]{hoangbinh_bb_numerical_rates}
\bibinfo{author}{\bibfnamefont{D.}~\bibnamefont{{Hoang-Binh}}},
  \bibinfo{journal}{\aap} \textbf{\bibinfo{volume}{238}}, \bibinfo{pages}{449}
  (\bibinfo{year}{1990}).

\bibitem[{\citenamefont{{Green} et~al.}(1957)\citenamefont{{Green}, {Rush}, and
  {Chandler}}}]{green_bb_numerical_rates}
\bibinfo{author}{\bibfnamefont{L.~C.} \bibnamefont{{Green}}},
  \bibinfo{author}{\bibfnamefont{P.~P.} \bibnamefont{{Rush}}},
  \bibnamefont{and} \bibinfo{author}{\bibfnamefont{C.~D.}
  \bibnamefont{{Chandler}}}, \bibinfo{journal}{\apjs}
  \textbf{\bibinfo{volume}{3}}, \bibinfo{pages}{37} (\bibinfo{year}{1957}).

\bibitem[{\citenamefont{{Goldwire}}(1968)}]{goldwire_bb_numerical_rates}
\bibinfo{author}{\bibfnamefont{H.~C.} \bibnamefont{{Goldwire}},
  \bibfnamefont{Jr.}}, \bibinfo{journal}{\apjs} \textbf{\bibinfo{volume}{17}},
  \bibinfo{pages}{445} (\bibinfo{year}{1968}).

\bibitem[{\citenamefont{{Burgess}}(1965)}]{burgess_bf_rates}
\bibinfo{author}{\bibfnamefont{A.}~\bibnamefont{{Burgess}}},
  \bibinfo{journal}{Mem. R. Astron. Soc.} \textbf{\bibinfo{volume}{69}},
  \bibinfo{pages}{1} (\bibinfo{year}{1965}).

\bibitem[{\citenamefont{{Gordon}}(1929)}]{gordon_formula}
\bibinfo{author}{\bibfnamefont{W.}~\bibnamefont{{Gordon}}},
  \bibinfo{journal}{Ann. Phys.} \textbf{\bibinfo{volume}{394}},
  \bibinfo{pages}{1031} (\bibinfo{year}{1929}).

\bibitem[{\citenamefont{{Brocklehurst}}(1971)}]{brocklehurst_rates_b}
\bibinfo{author}{\bibfnamefont{M.}~\bibnamefont{{Brocklehurst}}},
  \bibinfo{journal}{\mnras} \textbf{\bibinfo{volume}{153}},
  \bibinfo{pages}{471} (\bibinfo{year}{1971}).

\bibitem[{\citenamefont{Ueberhuber}(1997)}]{newton_cotes}
\bibinfo{author}{\bibfnamefont{C.}~\bibnamefont{Ueberhuber}},
  \emph{\bibinfo{title}{{Numerical Computation 2: Methods, Software, and
  Analysis}}} (\bibinfo{publisher}{Berlin: Springer}, \bibinfo{year}{1997}).

\bibitem[{\citenamefont{{Chluba} and
  {Sunyaev}}(2009{\natexlab{d}})}]{sunyaev_specdist}
\bibinfo{author}{\bibfnamefont{J.}~\bibnamefont{{Chluba}}} \bibnamefont{and}
  \bibinfo{author}{\bibfnamefont{R.~A.} \bibnamefont{{Sunyaev}}},
  \bibinfo{journal}{\aap} \textbf{\bibinfo{volume}{501}}, \bibinfo{pages}{29}
  (\bibinfo{year}{2009}{\natexlab{d}}), \eprint{arXiv:0803.3584}.

\bibitem[{\citenamefont{Padmanabhan and Finkbeiner}(2005)}]{paddy_decaydm_cmb}
\bibinfo{author}{\bibfnamefont{N.}~\bibnamefont{Padmanabhan}} \bibnamefont{and}
  \bibinfo{author}{\bibfnamefont{D.~P.} \bibnamefont{Finkbeiner}},
  \bibinfo{journal}{\prd} \textbf{\bibinfo{volume}{72}},
  \bibinfo{pages}{023508} (\bibinfo{year}{2005}),
  \eprint{arXiv:astro-ph/0503486}.

\bibitem[{\citenamefont{Chen and Kamionkowski}(2004)}]{chen_decaydm_cmb}
\bibinfo{author}{\bibfnamefont{X.-L.} \bibnamefont{Chen}} \bibnamefont{and}
  \bibinfo{author}{\bibfnamefont{M.}~\bibnamefont{Kamionkowski}},
  \bibinfo{journal}{\prd} \textbf{\bibinfo{volume}{70}},
  \bibinfo{pages}{043502} (\bibinfo{year}{2004}),
  \eprint{arXiv:astro-ph/0310473}.

\bibitem[{\citenamefont{{Wang} and {Li}}(2009)}]{spacetime_vary}
\bibinfo{author}{\bibfnamefont{X.}~\bibnamefont{{Wang}}} \bibnamefont{and}
  \bibinfo{author}{\bibfnamefont{M.}~\bibnamefont{{Li}}}
  (\bibinfo{year}{2009}), \eprint{arXiv:0904.1061}.

\bibitem[{\citenamefont{Slatyer et~al.}(2009)\citenamefont{Slatyer,
  Padmanabhan, and Finkbeiner}}]{slatyer_decaydm_cmb}
\bibinfo{author}{\bibfnamefont{T.~R.} \bibnamefont{Slatyer}},
  \bibinfo{author}{\bibfnamefont{N.}~\bibnamefont{Padmanabhan}},
  \bibnamefont{and} \bibinfo{author}{\bibfnamefont{D.~P.}
  \bibnamefont{Finkbeiner}} (\bibinfo{year}{2009}), \eprint{arXiv:0906.1197}.

\bibitem[{\citenamefont{Zahn and Zaldarriaga}(2003)}]{zahn_friedmann_rec}
\bibinfo{author}{\bibfnamefont{O.}~\bibnamefont{Zahn}} \bibnamefont{and}
  \bibinfo{author}{\bibfnamefont{M.}~\bibnamefont{Zaldarriaga}},
  \bibinfo{journal}{\prd} \textbf{\bibinfo{volume}{67}},
  \bibinfo{pages}{063002} (\bibinfo{year}{2003}),
  \eprint{arXiv:astro-ph/0212360}.

\bibitem[{\citenamefont{Anderson et~al.}(1999)}]{lapack}
\bibinfo{author}{\bibfnamefont{E.}~\bibnamefont{Anderson}}
  \bibnamefont{et~al.}, \emph{\bibinfo{title}{{LAPACK} Users' Guide}}
  (\bibinfo{publisher}{Society for Industrial and Applied Mathematics},
  \bibinfo{address}{Philadelphia, PA}, \bibinfo{year}{1999}),
  \bibinfo{edition}{3rd} ed., ISBN \bibinfo{isbn}{0-89871-447-8 (paperback)}.

\bibitem[{\citenamefont{{Press} et~al.}(1986)\citenamefont{{Press}, {Flannery},
  and {Teukolsky}}}]{nrbook}
\bibinfo{author}{\bibfnamefont{W.~H.} \bibnamefont{{Press}}},
  \bibinfo{author}{\bibfnamefont{B.~P.} \bibnamefont{{Flannery}}},
  \bibnamefont{and} \bibinfo{author}{\bibfnamefont{S.~A.}
  \bibnamefont{{Teukolsky}}}, \emph{\bibinfo{title}{{Numerical recipes. The art
  of scientific computing}}} (\bibinfo{publisher}{Cambridge: University Press,
  1986}, \bibinfo{year}{1986}).

\bibitem[{\citenamefont{Jitrik and Bunge}(2004)}]{jitrik_bb_quad_rates}
\bibinfo{author}{\bibfnamefont{O.}~\bibnamefont{Jitrik}} \bibnamefont{and}
  \bibinfo{author}{\bibfnamefont{C.~F.} \bibnamefont{Bunge}},
  \bibinfo{journal}{J. Phys. Chem. Ref. Data} \textbf{\bibinfo{volume}{33}},
  \bibinfo{pages}{1059} (\bibinfo{year}{2004}),
  \urlprefix\url{http://link.aip.org/link/?JPR/33/1059/1}.

\bibitem[{\citenamefont{{Johnson}}(2007)}]{johnson_quads}
\bibinfo{author}{\bibfnamefont{W.~R.} \bibnamefont{{Johnson}}},
  \emph{\bibinfo{title}{{Atomic Structure Theory}}}
  (\bibinfo{publisher}{Berlin: Springer}, \bibinfo{year}{2007}).

\bibitem[{\citenamefont{{Hey}}(2006)}]{hey_bb_rates}
\bibinfo{author}{\bibfnamefont{J.~D.} \bibnamefont{{Hey}}},
  \bibinfo{journal}{\jpb} \textbf{\bibinfo{volume}{39}}, \bibinfo{pages}{2641}
  (\bibinfo{year}{2006}).

\bibitem[{\citenamefont{{Strelnitski} et~al.}(1996)\citenamefont{{Strelnitski},
  {Ponomarev}, and {Smith}}}]{strel}
\bibinfo{author}{\bibfnamefont{V.~S.} \bibnamefont{{Strelnitski}}},
  \bibinfo{author}{\bibfnamefont{V.~O.} \bibnamefont{{Ponomarev}}},
  \bibnamefont{and} \bibinfo{author}{\bibfnamefont{H.~A.}
  \bibnamefont{{Smith}}}, \bibinfo{journal}{\apj}
  \textbf{\bibinfo{volume}{470}}, \bibinfo{pages}{1118} (\bibinfo{year}{1996}),
  \eprint{arXiv:astro-ph/9511118}.

\bibitem[{\citenamefont{{Vriens}}(1966)}]{clc}
\bibinfo{author}{\bibfnamefont{L.}~\bibnamefont{{Vriens}}},
  \bibinfo{journal}{Phys. Rev.} \textbf{\bibinfo{volume}{141}},
  \bibinfo{pages}{88} (\bibinfo{year}{1966}).

\bibitem[{\citenamefont{{Pengelly} and {Seaton}}(1964)}]{pengelly}
\bibinfo{author}{\bibfnamefont{R.~M.} \bibnamefont{{Pengelly}}}
  \bibnamefont{and} \bibinfo{author}{\bibfnamefont{M.~J.}
  \bibnamefont{{Seaton}}}, \bibinfo{journal}{\mnras}
  \textbf{\bibinfo{volume}{127}}, \bibinfo{pages}{165} (\bibinfo{year}{1964}).

\bibitem[{\citenamefont{{Spaans} and {Norman}}(1997)}]{spnorman}
\bibinfo{author}{\bibfnamefont{M.}~\bibnamefont{{Spaans}}} \bibnamefont{and}
  \bibinfo{author}{\bibfnamefont{C.~A.} \bibnamefont{{Norman}}},
  \bibinfo{journal}{\apj} \textbf{\bibinfo{volume}{488}}, \bibinfo{pages}{27}
  (\bibinfo{year}{1997}).

\bibitem[{\citenamefont{{Seljak} and {Zaldarriaga}}(1996)}]{cmbfast}
\bibinfo{author}{\bibfnamefont{U.}~\bibnamefont{{Seljak}}} \bibnamefont{and}
  \bibinfo{author}{\bibfnamefont{M.}~\bibnamefont{{Zaldarriaga}}},
  \bibinfo{journal}{\apj} \textbf{\bibinfo{volume}{469}}, \bibinfo{pages}{437}
  (\bibinfo{year}{1996}), \eprint{arXiv:astro-ph/9603033}.

\bibitem[{\citenamefont{{Kaulakys}}(1996)}]{kaulakys_wkb}
\bibinfo{author}{\bibfnamefont{B.}~\bibnamefont{{Kaulakys}}}
  (\bibinfo{year}{1996}), \eprint{arXiv:physics/9610018}.

\bibitem[{\citenamefont{Heim et~al.}(1989)\citenamefont{Heim, Trautmann, and
  Baur}}]{heim_bb_wkb_rates}
\bibinfo{author}{\bibfnamefont{T.~A.} \bibnamefont{Heim}},
  \bibinfo{author}{\bibfnamefont{D.}~\bibnamefont{Trautmann}},
  \bibnamefont{and} \bibinfo{author}{\bibfnamefont{G.}~\bibnamefont{Baur}},
  \bibinfo{journal}{\jpb} \textbf{\bibinfo{volume}{22}}, \bibinfo{pages}{727}
  (\bibinfo{year}{1989}),
  \urlprefix\url{http://stacks.iop.org/0953-4075/22/727}.

\bibitem[{\citenamefont{Delone et~al.}(1982)\citenamefont{Delone, Goreslavsky,
  and Krainov}}]{delone_bf_wkb_rates_a}
\bibinfo{author}{\bibfnamefont{N.~B.} \bibnamefont{Delone}},
  \bibinfo{author}{\bibfnamefont{S.~P.} \bibnamefont{Goreslavsky}},
  \bibnamefont{and} \bibinfo{author}{\bibfnamefont{V.~P.}
  \bibnamefont{Krainov}}, \bibinfo{journal}{\jpb}
  \textbf{\bibinfo{volume}{15}}, \bibinfo{pages}{L421} (\bibinfo{year}{1982}),
  \urlprefix\url{http://stacks.iop.org/0022-3700/15/L421}.

\end{thebibliography}
   \end{document}